%% file: main.tex
\documentclass{article}

\usepackage{abstract}
\usepackage{authblk}
\usepackage{algorithm}
\usepackage{algpseudocode}
\usepackage{float}
\usepackage[margin=1in]{geometry}
\usepackage{graphicx}
\usepackage[colorlinks,linkcolor=blue,citecolor=blue,urlcolor=blue]{hyperref}
\usepackage{natbib}
\usepackage{subfig}

\input{custom_commands}

\title{Temporal Functional Factor Analysis of Brain Connectivity}

\author{
    Kyle Stanley\thanks{Department of Biostatistics, Epidemiology, and Informatics, University of Pennsylvania. \texttt{kyle.stanley2@pennmedicine.upenn.edu}} \hspace{2em} 
    Nicole Lazar\thanks{Department of Statistics and Huck Institutes of the Life Sciences, Pennsylvania State University. \texttt{nfl5182@psu.edu}} \hspace{2em} 
    Matthew Reimherr\thanks{Department of Statistics, Pennsylvania State University, and Amazon Science. \texttt{mlr36@psu.edu}}
}

\date{}

\begin{document}

\maketitle

\begin{abstract}
Many analyses of functional magnetic resonance imaging (fMRI) examine functional connectivity (FC), or the statistical dependencies among distant brain regions. These analyses are typically exploratory, guiding future confirmatory research. In this work, we present an approach based on factor analysis (FA) that is well-suited to studying FC. FA is appealing in this context because its flexible model assumptions permit a guided investigation of its target subspace consistent with the exploratory role of connectivity analyses. However, applying FA to fMRI data poses three problems: (1) its target subspace captures short-range spatial dependencies that should be treated as noise, (2) it requires factorization of a massive spatial covariance, and (3) it overlooks temporal dependencies in the data. To address these limitations, we develop a factor model within the framework of functional data analysis--a field which views certain data as arising from smooth underlying curves. The proposed approach (1) uses matrix completion techniques to filter short-range spatial dependencies out of its target subspace, (2) employs a distributed algorithm for factorizing large-scale covariance matrices, and (3) leverages functional regression to exploit temporal dynamics. Together, these innovations yield a comprehensive and scalable method for studying FC.
\end{abstract}

\noindent \textbf{Keywords:} factor analysis, fMRI, functional connectivity, functional data analysis, matrix completion

\section{Introduction}
\label{sec:intro}

Functional magnetic resonance imaging (fMRI) is among the most widely used techniques for studying brain function. It is based on the haemodynamic response, a phenomenon wherein oxygenated blood rushes to active neuronal tissue. By measuring relative oxygenation levels across the brain, BOLD imaging allows researchers to indirectly study brain activity. During an fMRI session, a subject lies in a scanner that collects a temporal sequence of three-dimensional BOLD images, discretized by a grid of volume elements, called voxels \citep{lazar-2008,lindquist-2008}. The resulting data are both large and complex. Each brain image contains more than 100 thousand voxels, each session includes hundreds of these images, and many modern fMRI datasets contain hundreds of such sessions. Moreover, these data exhibit spatiotemporal dependencies that complicate their study. 

Analyses of fMRI data may be hypothesis- or data-driven. Hypothesis-driven studies aim to localize regions of the brain activated by a certain stimulus. In such studies, the researcher constructs a hypothesized time course for a voxel activated by the stimulus, then uses a statistical model to independently test whether each voxel's time course is related to the hypothesized one \citep{friston-etal-1995}. On the other hand, many data-driven studies seek to characterize \textit{functional connectivity (FC)}, or the statistical dependencies among distant brain regions \citep{friston-2011}. The simplest FC approach is to create a spatial map that visualizes correlations between the time course of a seed region and those of other regions \citep{biswal-etal-1995}. Other approaches, like principal component analysis (PCA; \citealp{jolliffe-1986}) and independent component analysis (ICA; \citealp{hyvarinen-etal-2004}), are based on linear latent variable models (LLVMs) that represent an fMRI scan as a set of spatial maps and their associated time courses. Data-driven approaches do not require specification of hypothetical time courses, making them particularly useful in the analysis of \textit{resting-state fMRI (rsfMRI)}, wherein subjects simply lie at rest in the scanner. Such methods play an exploratory role in brain science, aiding in the discovery of unanticipated connectivity that can inform hypothesis-driven confirmatory analyses. In this work, we develop a novel LLVM to study FC in resting-state data from the PIOP1 dataset of the Amsterdam Open MRI Collection (AOMIC; \citealp{snoek-et-al-2021}), which contain six-minute scans collected from 210 subjects at rest. 

\begin{figure}[!h]
    \centering
    \includegraphics[width=\linewidth]{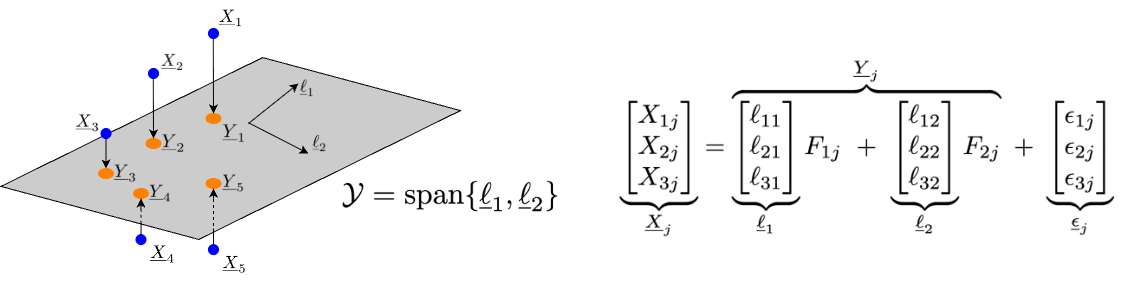}
    \caption{A depiction of the LLVM framework when $M = 3$, $J = 5$, and $K = 2$ (e.g., five time points of a three-voxel brain image modeled by a two-component LLVM).}
    \label{fig:intro-llvm}
\end{figure}

LLVMs describe an $M$-by-$J$ fMRI data matrix $\mb{X}$ ($M$ the number of voxels and $J$ the number of time points) as the sum of a ``signal'' term $\mb{Y}$ having rank $K < \min\{M,J\}$ and a ``noise'' term $\mb{\epsilon}$:
\begin{equation*}
    \mb{X} = \underbrace{\mb{L} \mb{F}}_{\mb{Y}} + \mb{\epsilon},
\end{equation*}
where the columns of $\mb{L} \in \mbb{R}^{M \times K}$ contain $K$ spatial maps and the rows of $\mb{F} \in \mbb{R}^{K \times J}$ contain their accompanying time courses. Simply put, LLVMs posit that the interesting behavior in an fMRI scan can be described by a few spatial maps that are activated according to their time courses. See Figure \ref{fig:intro-llvm} for an illustration of this LLVM framework. Choosing from among the many possible $(\mb{L}, \mb{F})$ reduces to answering the following questions: 
\begin{enumerate}
    \item What is the ``target subspace'' (fixes $\mcal{Y} = \text{span}(\mb{L})$)?
    \item How should the target subspace be ``expressed'' (fixes $\mb{L}$)?
\end{enumerate}
In what follows, we critique how existing LLVMs answer these questions, and offer better calibrated responses that motivate development of our LLVM.

Fixing the target subspace is equivalent to specifying how noise contaminates the spatial covariance. To see this, let $\mb{C} = \mb{X}\mb{X}^T = \mb{C}_Y + \mb{C}_\epsilon$ be the decomposition of the spatial covariance into signal and noise covariances. Given the noise covariance $\mb{C}_\epsilon$, the signal covariance $\mb{C}_Y$ can be written as $\mb{L}\mb{L}^T$ for some $M$-by-$K$ matrix $\mb{L}$, thereby fixing $\mcal{Y} = \text{span}(\mb{L})$. Noiseless PCA and ICA set $\mb{C}_\epsilon$ equal to the zero matrix (i.e., all variation is viewed as signal) while their noisy analogues \citep{tipping-bishop-1999,beckmann-smith-2004} set $\mb{C}_\epsilon = \sigma^2\mb{I}$ for some $\sigma > 0$ (i.e., all variation except ``background noise'' is viewed as signal). This raises the question: is the noise structure specified by these models well-calibrated to the study of FC? Since it is unreasonable to expect the low-rank signal term of an LLVM to span variation at all spatial scales, a more suitable LLVM should absorb short-range spatial dependencies. Classifying such structure as noise is especially sensible in FC, where the aim is to describe long-range spatial dependencies.

Fixing $\mcal{Y}$ does not uniquely identify an LLVM, as there exist numerous expressions of $\mcal{Y}$: $\text{span}(\mb{L}\mb{T}) = \text{span}(\mb{L}) = \mcal{Y}$ for any invertible $K$-by-$K$ matrix $\mb{T}$. Many LLVMs fix $\mb{L}$ by imposing additional constraints. For instance, PCA assumes the columns of $\mb{L}$ are orthogonal and the rows of $\mb{F}$ are uncorrelated, while ICA assumes the rows of $\mb{F}$ are statistically independent. This leads to the question: in the study of FC, is it appropriate to limit oneself to a single expression of the target subspace? Given the exploratory nature of these studies, a more apt approach would permit a principled investigation of the target subspace, and might lead to the discovery of connectivity structures obscured by the unique representations of other LLVMs. Factor analysis (FA; \citealp{harman-1976,johnson-wichern-2002}), a method rarely encountered in FC research, provides a natural framework in which to conduct subspace exploration, and serves as the starting point in development of our approach. 

FA posits that an observation vector $\und{X} \in \mbb{R}^M$ (e.g., BOLD observations of an $M$-voxel brain image at a fixed time point) having covariance $\mb{C} \in \mbb{R}^{M\times M}$ is the sum of a \textit{common} component, given by linear combinations of zero-mean factors $\und{F} \in \mbb{R}^{K}$ having covariance $\mbb{E}[\und{F}\und{F}^T] = \mb{I}$, and a \textit{unique} component, given by a vector of uncorrelated zero-mean errors $\und{\epsilon} \in \mbb{R}^M$:
\begin{equation*}
    \und{X} = \mb{L} \und{F} + \und{\epsilon},
\end{equation*}
where $\mb{L}\in \mbb{R}^{M \times K}$ is called the \textit{loading matrix}. If $\mb{D} \in \mbb{R}^{M\times M}$ denotes the diagonal covariance matrix of $\und{\epsilon}$, then $\mb{C}$ decomposes as
\begin{equation}
\label{eqn:fa-cov}
    \mb{C} = \mb{L}\mb{L}^T + \mb{D}.
\end{equation}
Since $\mb{D}$ is diagonal, $\mb{L}\mb{L}^T$ must account for all variation off the diagonal of $\mb{C}$. This observation sheds light on the named components of the model: the common component describes all between-variable variation while the unique component describes leftover within-variable variation. After fixing $\mcal{Y} = \text{span}(\mb{L})$, FA explores this target subspace through through various \textit{factor rotations}. In \textit{orthogonal} rotation, one rotates the initial expression by an orthonormal matrix $\mb{R} \in \mbb{R}^{K \times K}$ to preserve $\mcal{Y}$ and factor uncorrelatedness:
\begin{align*}
    \mb{L}^* \und{F}^* & = (\mb{L}\mb{R}) (\mb{R}^T \und{F}) = \mb{L}\und{F} \\
    \mbb{E}[\und{F}^*\und{F}^{*T}] & = \mb{R}^T \mbb{E}[\und{F}\und{F}^{T}] \mb{R} = \mb{I}.
\end{align*}
In \textit{oblique} rotation, one rotates the initial expression by an invertible matrix $\mb{T} \in \mbb{R}^{K \times K}$ to preserve $\mcal{Y}$ and induce correlations among the factors:
\begin{align*}
    \mb{L}^* \und{F}^* & = (\mb{L}\mb{T}^{-1}) (\mb{T} \und{F}) = \mb{L}\und{F} \\
    \mbb{E}[\und{F}^*\und{F}^{*T}] & = \mb{T}^T \mbb{E}[\und{F}\und{F}^{T}] \mb{T} = \mb{T} \mb{T}^T.
\end{align*}
The matrices $\mb{R}$ and $\mb{T}$ are not chosen arbitrarily. Typically, one selects $\mb{R}$ or $\mb{T}$ to optimize some objective criterion--like varimax \citep{kaiser-1958} or quartimin \citep{jennrich-sampson-1966}--that encourages a more interpretable \textit{simple} loading structure, in which (loosely speaking) $\mb{L}$ contains a few high-magnitude weights and many near-zero weights. In this way, factor rotation enables a principled investigation of $\mcal{Y}$ that lends itself to the inherently exploratory task of studying FC.  

There are, however, three obstacles impeding application of FA to fMRI. First, the unique component fails to capture short-range spatial dependencies. As discussed, these dependencies ought to reside within the noise term since (a) such structure arises naturally in the residuals of low-rank LLVMs fit to fMRI data, and (b) FC is concerned only with long-range spatial dependencies. Second, model estimation requires factorization of a massive (voxel-by-voxel) spatial covariance. Other LLVMs, like PCA and ICA, circumvent this computational hurdle by transposing the problem, replacing $\mb{X}$ with $\mb{X}^T$, so that estimation is based on a much smaller temporal covariance \citep{mckeown-etal-1998,beckmann-smith-2004}. This strategy, however, prevents the decoupling of long- and short-range spatial variation, and is thus ineffectual for a model that addresses the first obstacle. Third, FA overlooks the temporal structure in fMRI data, treating the brain images $\und{X}_1, \dots, \und{X}_J$ of an fMRI session as independent. A more appropriate approach would exploit this structure to improve estimation of temporal quantities. 

To develop a better-suited factor model, we place fMRI within the framework of functional data analysis (FDA; \citealp{ramsay-silverman-2005}, \citealp{wang-et-al-2016}, \citealp{kokoszka-reimherr-2017}). Data are functional if it is natural to assume they arise from some smooth underlying curve. Since one may treat BOLD observations of an fMRI scan as noisy realizations from some smooth function over a four-dimensional spatiotemporal domain, fMRI (and other spatiotemporal) data are an example of multidimensional functional data. Though used sparingly by the neuroimaging community, FDA tools occasionally appear in the literature: functional PCA (FPCA, \citealp{viviani-etal-2005}), functional regression \citep{reiss-etal-2015, reiss-etal-2017}, and functional graphical models \citep{li-solea-2018}. To tailor FA to fMRI data, we extend the FA model to this spatiotemporal functional setting.

This paper builds upon the work of and may be viewed as a sequel to \cite{stanley-etal-2025}. Their key insight was to extend FA to discretely-observed functional data by replacing the common-unique decomposition of multivariate FA with a global-local decomposition that separates long- and short-range spatial dependencies. Leveraging the work of \cite{descary-panaretos-2019}, they show that if the global covariance is smooth with finite rank and the local covariance is banded, then this decomposition is identifiable. However, their approach, called functional FA (FFA), still requires factorization of a large spatial covariance and ignores the temporal dependencies in fMRI data. They accommodate these limitations by restricting analysis to a horizontal slice of the brain and prewhitening voxel time courses. This work addresses the shortcomings of  \cite{stanley-etal-2025}. First, it leverages the distributed algorithm of \cite{stanley-etal-2025+} to enable full-brain model estimation. Second, it acknowledges the temporal structure of fMRI data via incorporation of temporal factors whose scores may be computed using techniques from functional regression. The resulting approach, called \textit{temporal FFA (TFFA)}, represents a collection of fMRI scans as a set of common spatial maps and their affiliated subject-wise time courses. 

We begin, in Section \ref{sec:model}, by presenting the TFFA model at both infinite- and finite-resolution. Section \ref{sec:est} develops estimation methods for the finite-resolution model, which are studied in the simulations of Section \ref{sec:sim}. We then, in Section \ref{sec:aomic}, use TFFA to analyze the AOMIC resting-state data. We conclude with some final remarks in Section \ref{sec:disc}.

\section{The Model and its Identification}
\label{sec:model}

Since there are four dimensions in fMRI, three spatial and one temporal, there are actually several ways to view such data as functional--e.g., a collection of (i) spatial functions indexed over time, (ii) temporal functions indexed over space, or (iii) spatiotemporal functions indexed (in the multi-subject setting) over subjects. In this paper, we adopt view (iii).

We present our factor model for spatiotemporal functional data in two data observation paradigms, the first serving as a stepping stone to the second. The first assumes data are observed completely (e.g., a brain scan with infinite spatiotemporal resolution, making it a function), while the second assumes data are observed discretely on a grid (e.g., a brain scan with finite spatiotemporal resolution, making it a tensor). Unless coerced into some non-traditional format, fMRI data falls into this latter paradigm.

\subsection{Infinite Resolution}
\label{sec:model-ir}

Let $X_i$ be the $i$th subject's zero-mean spatiotemporal random function in $L^2([0,1]^{D+1})$, which is endowed with the typical inner product and norm:
\begin{align*}
    \ip{f}{g}_{L^2} = \int_{[0,1]^{D+1}} f(\und{s}, t) g(\und{s}, t) d\und{s}dt \hspace{1em} \text{and} \hspace{1em} \norm{f}_{L^2}^2 = \ip{f}{f}_{L^2}. 
\end{align*}
In fMRI, $D=3$ and $X_i(\und{s},t)$ is a scan whose BOLD observations are accessible at infinite spatial and temporal resolution. The covariance function of $X_i$ is given by 
\begin{align*}
    c_i(\und{s}, t, \und{s}', t') = \mbb{E}[X_i(\und{s}, t)X_i(\und{s}', t')].
\end{align*}
We say $X_i$ follows a \textit{temporal functional factor model (TFFM)} with $K$ factors if 
\begin{align*}
    X_i(\und{s}, t) = \underbrace{\sum_{k=1}^K l_k(\und{s})F_{i,k}(t)}_{Y_i(\und{s}, t)} + \epsilon_i (\und{s}, t),
\end{align*}
where $Y_i$ and $\epsilon_i$ are the \textit{global} and \textit{local} components and, within the global component, $l_k$ and $F_{i,k}$ are the $k$th \textit{loading function} and \textit{functional factor}. The model assumes the loadings $l_k$ are sufficiently smooth (in a sense to be defined in Section \ref{sec:model-ir-id}), and that the local covariance is $\und{\delta}$-banded when integrated over time, i.e., $\int\mbb{E}[\epsilon_i(\und{s}, t)\epsilon_i(\und{s}', t)]dt = 0$ when $\abs{s_d - s'_d} \geq \delta_d > 0$ for some $d = 1, \dots, D$. We call $Y_i$ the global term since, as a finite separable sum with smooth spatial components, it contains all spatial variation in $X_i$ that is sustained and long-range. On the other hand, we call $\epsilon_i$ the local term since its banded spatial covariance ensures it absorbs all sustained short-range spatial variation in the data that is not incidentally captured by $Y_i$. That the two terms characterize ``sustained'' variation arises from the integral assumption on the local covariance (see Remark \ref{rk:int-pw} for details). In what follows, we often condense notation by writing $\und{l}(\und{s}) = [l_1(\und{s}), \dots, l_K(\und{s})]$ and $\und{F}(t) = [F_1(t), \dots, F_K(t)]$. 

Let $g_i$ and $b_i$ denote the covariance functions of $Y_i$ and $\epsilon_i$, respectively. Assume the random subject-specific factors $F_{i,k}$ and local term $\epsilon_i$ are mean-zero and satisfy
\begin{align}
    \int \mbb{E}[\und{F}_i(t) \und{F}_i^T(t)] dt & = \mb{I} \ \text{and} \label{eqn:ir-assump-facs-or} \\
    \int \mbb{E}[\und{F}_i(t)\epsilon_i(\und{s},t)] dt & = \und{0}. \label{eqn:ir-assump-facs-err}
\end{align}
That is, the factors are uncorrelated with unit variance ``on average'', and the factors and errors are uncorrelated ``on average''. If these conditions hold, then the marginal spatial covariance $c_i^s$, obtained by integrating $c_i(\und{s},t,\und{s}',t)$ over $t$, decomposes as
\begin{equation}
\label{eqn:ir-cov-space}
    c_i^s(\und{s}, \und{s}') := \int_0^1 c_i(\und{s},t,\und{s}',t) dt = \underbrace{\sum_{k=1}^K l_k(\und{s}) l_k(\und{s}')}_{g^s(\und{s}, \und{s}')} + b_i^s(\und{s}, \und{s}),
\end{equation}
where $g^s$ is the global spatial covariance common to all subjects, and $b_i^s$ is the local spatial covariance for subject $i$. Note that, in the decomposition of (\ref{eqn:ir-cov-space}), only $g^s$ does not depend on the subject. It turns out that the local spatial covariance need not be the same for every subject in order to identify $g^s$. By fixing $g^s$, TFFA fixes its target subspace $\mcal{Y} = \text{span}(l_1, \dots, l_K)$.

If $\lambda_k$ and $\eta_k$ are the $k$th eigenvalue and eigenfunction of $g^s$, respectively, it is easy to show that one may rotate the scaled eigenfunctions $\Tilde{\eta}_k = \lambda^{1/2} \eta_k$ by some orthonormal matrix $\mb{R}$ to recover the $l_k$ (see Section \ref{sec:supp-ti-or} of the supplement). However, one does not know this rotation in practice. In fact, \textit{every} rotation of the scaled eigenfunctions leads to a statistically equivalent model (i.e., each produces the global term and uncorrelated factors; see Section \ref{sec:supp-ti-or} of the supplement). In practice, we exploit this ambiguity by exploring the global subspace $\mcal{Y}$ through simplicity-inducing orthogonal rotation procedures, which we review in Section \ref{sec:est-subexp}.

\begin{remark}[Integrated vs. Pointwise Conditions]
\label{rk:int-pw}
    To guarantee the spatial covariance decomposition of (\ref{eqn:ir-cov-space}), the TFFM imposes several conditions which involve integrating over time. A stronger set of assumptions leading to the same decomposition replaces these integral conditions with pointwise conditions (e.g., factors are pointwise uncorrelated with unit variance). Even in the resting-state setting, wherein the absence of stimuli might lead one to expect temporal uniformity, these pointwise assumptions are too restrictive. By instead constraining the integrals, we impose conditions on ``average'' behavior rather than pointwise behavior. This provides a great deal of flexibility. For instance, factors may be transiently correlated and have time-varying variance so long as they are uncorrelated with unit variance ``on average''. This averaging does, however, admit a pathological peculiarity. Consider two distant spatial locations $\und{s}$ and $\und{s}'$ whose BOLD observations have covariance $a$ when $0 \leq t < 1/2$ and $-a$ when $1/2 \leq t \leq 1$. Though this contrived data contains pointwise global variation (i.e., $g(\und{s}, t, \und{s}', t) \neq 0$ for all $t$), integration removes this structure from the global spatial covariance (i.e., $g^s(\und{s}, \und{s}') = 0$). Consequently, the TFFM may fail to detect transient global structure and is best understood as decomposing spatial variation into ``sustained'' global and local modes.   
\end{remark}

\subsubsection{Correlated Factors}
\label{sec:model-ir-ob}

We refer to the model thus presented as the \textit{orthogonal} TFFM, as its factors are (on average) uncorrelated. The \textit{oblique} TFFM relaxes the assumption in (\ref{eqn:ir-assump-facs-or}), permitting sustained correlation among the factors:
\begin{equation}
\label{eqn:ir-assump-facs-ob}
    \int \mbb{E}[\und{F}_i(t) \und{F}_i^T(t)] dt = \mb{H},
\end{equation}
where $\mb{H}$ is some unknown covariance matrix constrained by $\text{diag}(\mb{H}) = \mb{I}$ to fix the scale of factor variances. Under (\ref{eqn:ir-assump-facs-ob}), the global component in the spatial covariance decomposition of (\ref{eqn:ir-cov-space}) admits a decomposition different from that  in the orthogonal model: 
\begin{equation*}
    g^s(\und{s},\und{s}') = \sum_{k=1}^K \Tilde{l}_k(\und{s}) \Tilde{l}_k(\und{s}'),
\end{equation*}
where $\und{\Tilde{l}}(\und{s}) = \und{l}(\und{s}) \mb{V} \mb{\Lambda}^{1/2}$ and $\mb{H} = \mb{V} \mb{\Lambda} \mb{V}^T$ is the eigendecomposition of $\mb{H}$. As we show in Section \ref{sec:supp-ti-ob} of the supplement, one may recover the loadings $l_k$ by applying some unknown invertible transformation $\mb{T}$ to the scaled eigenfunctions $\Tilde{\eta}_k$. The oblique model is thus even more ambiguous than the orthogonal one. Rather than fixing $\mb{R}$ within the space of orthonormal matrices, we must now fix $\mb{T}$ within the much larger space of unit-diagonal invertible matrices. Though a theoretical inconvenience, this added indeterminacy has practical appeal: by expanding the space of transformations, one may discover simpler expressions of the spatial loadings than is possible in the orthogonal setting. This is accomplished via a suite of oblique rotation procedures, which we review in Section \ref{sec:est-subexp}. Since the sole difference between the orthogonal and oblique models is the type of transformation applied to the scaled eigenfunctions we present our work in the orthogonal setting, highlighting any deviations that arise in the oblique setting.

\subsubsection{Global-Local Uniqueness}
\label{sec:model-ir-id}

Regardless of whether one assumes orthogonal or oblique factors, recovery of the loading functions requires knowledge of $g^s$. It is thus critical that $g^s$ be uniquely determined by $c_i^s$. The assumptions of the TFFM impose two conditions on the spatial covariance decomposition: $g^s$ has finite rank, and $b_i^s$ is $\und{\delta}$-banded. These restrictions, however, do not sufficiently constrain the problem. Given $c_i^s$, we know $g^s$ off the band, but it is not clear how to uniquely extend $g^s$ onto the band. It turns out that a particular formulation of smoothness, called \textit{real analyticity}, permits this unique extension. Guided by the work of \cite{descary-panaretos-2019}, Theorem 1 from Section 3.1 of the supplement in \cite{stanley-etal-2025} shows that if, in addition, the $l_k$ are real analytic, then $g^s$ is uniquely determined by $c_i^s$. The proof exploits the so-called continuation property of analytic functions, which states that if a function is analytic over some domain, but is known only on an open subset of that domain, then the function extends uniquely to the rest of the domain.

\subsection{Finite Resolution}
\label{sec:model-fr}

In practice, we observe a spatiotemporal function $X_i$ at finite spatial and temporal resolution. Let $\{ I_{\und{m}} \}_{\und{m} \in \mbb{N}_{\und{M}}}$ be the partition of $[0,1]^D$ into cells of size $M_1^{-1} \times \dots \times M_D^{-1}$, where $\und{M} = (M_1, \dots, M_D)$ is the spatial resolution and $\mbb{N}_{\und{M}} = \mbb{N}_{M_1}\times \dots \times \mbb{N}_{M_D}$ is the set of all $D$-dimensional multi-indices with $d$th dimension no greater than $M_d$ (i.e., $\mbb{N}_{M_d} = \{ 1, \dots , M_d \}$). Given that fMRI partitions three-dimensional space into a uniform grid of voxels, one can view $I_{\und{m}}$ as representing the voxel at grid coordinate $\und{m} = (m_1, m_2, m_3)$. We assume, as is true in multi-subject fMRI data, that spatial images are comprised of the same $M = M_1 \dots M_D$ discrete points $\mcal{S} = \{ \und{s}_{\und{m}} \in I_{\und{m}} : \und{m} \in \mbb{N}_{\und{M}} \}$ across subjects and time points. We also assume that each $X_i$ is observed on the same evenly-spaced $J$-resolution temporal grid $\mcal{T} = \{ t_j : j \in \mbb{N}_J \}$. As detailed in Remark \ref{rk:int-approx}, we informally require this temporal grid to be sufficiently dense relative to the temporal smoothness of variation in the data. This scheme gives rise to the following observation model: 
\begin{equation*}
    X_i (\und{s}_{\und{m}}, t_j) = \underbrace{\sum_{k=1}^K l_k(\und{s}_{\und{m}})F_{i,k}(t_j)}_{Y_i(\und{s}_{\und{m}}, t_j)} + \epsilon_i (\und{s}_{\und{m}}, t_j), \ \und{m} \in \mbb{N}_{\und{M}}, \ j \in \mbb{N}_J.
\end{equation*}
Summarizing the functional terms in the model with the tensors 
\begin{align*}
    \mcal{X}_i (\und{m}, j) = X_i(\und{s}_{\und{m}}, t_j), \hspace{2em} \mcal{Y}_i (\und{m}, j) = Y_i(\und{s}_{\und{m}}, t_j), \hspace{2em} \mcal{E}_i (\und{m}, j) = \epsilon_i(\und{s}_{\und{m}}, t_j),
\end{align*}
\vspace{-1.75em}
\begin{align*}
    \mcal{L}_k(\und{m}) = l_k(\und{s}_{\und{m}}), \hspace{2em} \mcal{F}_{i,k}(j) = F_{i,k}(t_j),
\end{align*}
we can compactly write the \textit{$(\und{M}, J)$-resolution TFFM} with $K$ factors as
\begin{align*}
    \mcal{X}_i = \underbrace{\sum_{k=1}^K \mcal{L}_k \otimes \mcal{F}_{i,k}}_{\mcal{Y}_i} + \mcal{E}_i,
\end{align*}
where $\otimes$ denotes the usual tensor product.

\begin{remark}(Integral Approximations)
\label{rk:int-approx}
    At finite resolution, we informally assume that $\mcal{T}$ is dense relative to the smoothness of several covariances: $\mbb{E}[\epsilon_i(\und{s},t) \epsilon_i(\und{s}',t)]$, $\mbb{E}[F_{i,k}(t) F_{i,k'}(t)]$, and $\mbb{E}[F_{i,k}(t) \epsilon_i(\und{s},t)]$. This ensures discrete approximations to the integral conditions of Section \ref{sec:model-ir} do not incur too much error: 
    \begin{align*}
        \int \mbb{E}[\epsilon_i(\und{s},t) \epsilon_i(\und{s}',t)]dt & \approx J^{-1} \sum_j \mbb{E}[\epsilon_i(\und{s},t_j) \epsilon_i(\und{s}',t_j')], \\
        \int \mbb{E}[F_{i,k}(t) F_{i,k'}(t)]dt & \approx J^{-1} \sum_j \mbb{E}[F_{i,k}(t_j) F_{i,k'}(t_j)], \text{ and} \\
        \int \mbb{E}[F_{i,k}(t) \epsilon_i(\und{s},t)]dt & \approx J^{-1} \sum_j \mbb{E}[F_{i,k}(t_j) \epsilon_i(\und{s},t_j)]. 
    \end{align*}
    The decomposition of the spatial covariance in (\ref{eqn:fr-scov-decomp}) depends on these high-fidelity approximations. Note that even if functional factors and error functions are themselves temporally rough, it is reasonable to assume that expectations among and between these functions are smooth. 
\end{remark}

Now, use the following $M_1 \times \dots \times M_D \times J \times M_1 \times \dots \times M_D \times J$ tensors to summarize discretizations of the model's covariance functions: 
\begin{align*}
    \mcal{C}_i (\und{m}, j, \und{m}', j') & = c_i (\und{s}_{\und{m}}, t_j, \und{s}_{\und{m}'}, t_{j'}), \\
    \mcal{G}_i (\und{m}, j, \und{m}', j') & = g_i (\und{s}_{\und{m}}, t_j, \und{s}_{\und{m}'}, t_{j'}), \\
    \mcal{B}_i (\und{m}, j, \und{m}', j') & = b_i (\und{s}_{\und{m}}, t_j, \und{s}_{\und{m}'}, t_{j'}). \\
\end{align*}
Then the full covariance decomposes as
\begin{equation}
\label{eqn:fr-cov-decomp}
\mcal{C}_i = \mcal{G}_i + \mcal{B}_i.
\end{equation}
In the orthogonal context, summing each tensor in (\ref{eqn:fr-cov-decomp}) over $j$ yields the following decomposition of the spatial covariance:
\begin{equation}
\label{eqn:fr-scov-decomp}
    \mcal{C}_i^s = \underbrace{\sum_{k=1}^K \mcal{L}_k \otimes \mcal{L}_k}_{\mcal{G}^s} + \mcal{B}_i^s.
\end{equation}
Similar to the infinite-resolution setting, the $\mcal{L}_k$ are equal to the scaled eigentensors, $\Tilde{\mcal{H}} = \lambda^{1/2}\mcal{H}_k$, of $\mcal{G}^s$ up to an orthogonal rotation, where $\lambda_k$ and $\mcal{H}_k$ are the $k$th eigenvalue and eigentensor of $\mcal{G}^s$, respectively. Note also that since $b_i$ is $\und{\delta}$-banded in space, the tensor $\mcal{B}_i^s$ is banded. Moving forward, we declutter notation by replacing $M_1 \times \dots \times M_D$ with $\mfrak{M}$ where appropriate.

\subsubsection{Identifiability}
\label{sec:model-fr-id}

Analogous to the problem posed at infinite resolution, we must uniquely identify $\mcal{G}^s$ from $\mcal{C}_i^s$. This, however, is not obviously possible since analyticity, used to establish uniqueness at infinite resolution, is a property of functions, not of tensors. It turns out that $\mcal{G}^s$ may still be identified from $\mcal{C}_i^s$ and that analyticity plays a crucial role. \cite{stanley-etal-2025}, guided by \cite{descary-panaretos-2019}, establish that $\mcal{G}^s$ is unique provided each $\delta_d$ is less than 1/2 and
\begin{equation}
\label{eqn:fr-assump-params}
K \leq K^* = \prod_{d=1}^D \left\lfloor (1/2 - \delta_d)M_d - 1 \right\rfloor.
\end{equation}

\section{Estimation}
\label{sec:est}

Estimation of quantities in the finite-resolution model proceeds in four stages: (i) subspace estimation, (ii) subspace exploration, (iii) postprocessing, and (iv) factor score estimation. In (i), we use low-rank covariance matrix completion to estimate the global spatial covariance and define an initial expression of the associated subspace using its scaled eigentensors. In (ii), we explore this global subspace via various orthogonal and oblique rotations that encourage a structure simpler than that of the eigentensors. In (iii), we sharpen our estimate of the global subspace and make its expression more interpretable through smoothing and shrinkage. Finally, in (iv), we treat the postprocessed loadings of (iii) as fixed within the TFFM then estimate scores for the random factors via function-on-scalar regression.

\subsection{Subspace Estimation}
\label{sec:est-subest}

Recall that when the data are observed at resolution $(\und{M}, J)$ the spatial covariance for subject $i$ decomposes as
\begin{equation*}
    \mcal{C}_i^s = \underbrace{\sum_{k=1}^K \mcal{L}_k \otimes \mcal{L}_k}_{\mcal{G}^s} + \mcal{B}_i^s.
\end{equation*}
Averaging the subject-wise spatial covariances produces a structurally equivalent (smooth and low-rank, plus banded) decomposition of the \textit{average spatial covariance},
\begin{equation*}
\mcal{C}^s := \frac{1}{n} \sum_{i=1}^n \mcal{C}_i^s = \sum_{k=1}^K \mcal{L}_k \otimes \mcal{L}_k + \mcal{B}^s,
\end{equation*}
where $\mcal{B}^s = n^{-1}\sum_i \mcal{B}_i^s$. Given a finite-resolution sample of spatiotemporal functions summarized by the tensors $\mcal{X}_1, \dots, \mcal{X}_n \in \mbb{R}^{\mfrak{M}\times J}$, our goal is to estimate the global spatial covariance $\mcal{G}^s$ from the \textit{empirical average spatial covariance},
\begin{equation*}
    \hat{\mcal{C}}^s = \frac{1}{nJ} \sum_{i=1}^n \sum_{j=1}^J (\mcal{X}_i - \bar{\mcal{X}})_j \otimes (\mcal{X}_i - \bar{\mcal{X}})_j,
\end{equation*}
where $\bar{\mcal{X}} = n^{-1} \sum_{i} \mcal{X}_i$, and $(\mcal{X}_i - \bar{\mcal{X}})_j$ denotes the $j$th time point of the $i$th subject's centered observation. 

Our definition and implementation of $\hat{\mcal{G}}^s$ is similar to that of the global covariance estimator in the FFM of \cite{stanley-etal-2025}. The estimator $\hat{\mcal{G}}^s$ is a low-rank covariance tensor that is a good approximation to $\hat{\mcal{C}}^s$ off the band: 
\begin{equation}
\label{eqn:est-g-fr}
\hat{\mcal{G}}^s = \underset{\theta \in 
 \Theta_{\und{M}}}{\arg\min} \left\{ \norm{\mcal{Z} \circ \left( \hat{\mcal{C}}^s - \theta \right)}_F^2 \right\} + \tau \text{rank}(\theta),
\end{equation}
where $\Theta_{\und{M}}$ is the space of $\mfrak{M} \times \mfrak{M}$ covariance tensors with trace norm bounded by that of $\hat{\mcal{C}}^s$, $\mcal{Z}$ is the $\mfrak{M} \times \mfrak{M}$ ``band-deleting'' tensor defined by $\mcal{Z}(\und{m}, \und{m}') = \mathds{1}\left\{ \abs{m_d - m_d'} > \lceil M_d / 4 \rceil, \text{ for all } d = 1, \dots, D \right\}$, and $\tau > 0$ is a rank parameter. The estimator for $\hat{\mcal{G}}^s$ may then be used to define estimators for the number of factors $K$ and initial loadings $\mcal{L}_k$: 
\begin{equation*}
    \hat{K} = \text{rank}(\hat{\mcal{G}}^s) \hspace{1em} \text{and} \hspace{1em} \hat{\mcal{L}}_k = \hat{\lambda}_k^{1/2} \hat{\mcal{H}}_k,
\end{equation*}
for $k = 1, \dots, K$, where $\hat{\lambda}_k$ and $\hat{\mcal{H}}_k$ are the $k$th eigenvalue and eigentensor of $\hat{\mcal{G}}^s$, respectively. \cite{stanley-etal-2025} encourage smoothness in their initial loading estimates by including a roughness penalty in their objective function. To achieve smoothness, we instead recommend smoothing the loadings as a postprocessing step (see Section \ref{sec:est-pp-smooth}). Though less streamlined than the penalization strategy of \cite{stanley-etal-2025}, postprocessed smoothing allows for finer control of loading-specific smoothness since each tensor is smoothed independently. 

To describe how one performs the optimization in (\ref{eqn:est-g-fr}), let us first assume that we have already chosen a rank parameter $\tau^*$. Define the functional $f: \mbb{R}^{\mfrak{M}\times \mfrak{M}} \to \mbb{R}$ by $f(\theta) = \norm{ \mcal{Z} \circ \left(\hat{\mcal{C}}^s - \theta \right) }_F^2$. We then estimate $\mcal{G}^s$ as follows: 

\begin{enumerate}
    \item[(a)] Solve the optimization problem 
    \begin{align*}
        \min_{0 \preceq \theta \in \mbb{R}^{\mfrak{M}\times\mfrak{M}}} f(\theta) \hspace{1em} \text{subject to} \hspace{1em} \text{rank}(\theta) \leq j,
    \end{align*}
    for $j \in \{ 1, \dots, K^* \}$, obtaining minimizers $\hat{\theta}_1, \dots, \hat{\theta}_{K^*}$.
    \item[(b)] Compute the quantities $\{ f(\hat{\theta}_j) + \tau^* j: j = 1, \dots , K^* \}$ and determine the $j$ furnishing the minimum quantity. Declare the corresponding $\hat{\theta}_j$ to be the estimator $\hat{\mcal{G}}^s$.
\end{enumerate}

In practice, we do not explicitly choose the rank parameter $\tau$. Since each $\tau$ implies a choice of rank $j_\tau$, and thus some $f(\hat{\theta}_{j_\tau})$, we use the nonincreasing function $j \mapsto f(\hat{\theta}_j)$ to implicitly select $\tau$. Specifically, we find the smallest $j$ such that $f(\hat{\theta}_j) < c$ for some threshold $c$, then set $\hat{\mcal{G}^s}$ to the corresponding minimizer. The threshold $c$ should be small enough for $f$ evaluated at the estimator to be low, but not so small that the estimator has large rank. We thus select the rank of the estimator by identifying the elbow of the scree-type plot generated by the function $j \mapsto f(\hat{\theta}_j)$.

\begin{figure}[!h]
\centering
\subfloat[Small $\tau$]{
    \label{fig:scree-illust-1}\includegraphics[width=0.8\linewidth]{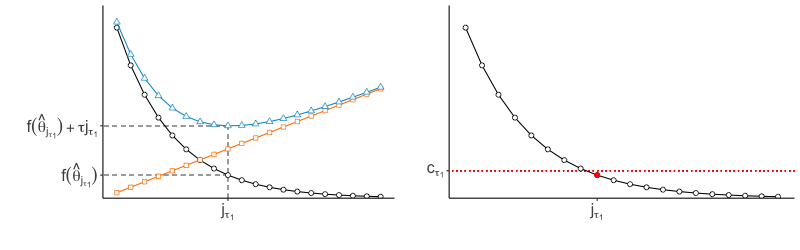}
}\\
\subfloat[Large $\tau$]{
    \label{fig:scree-illust-2}\includegraphics[width=0.8\linewidth]{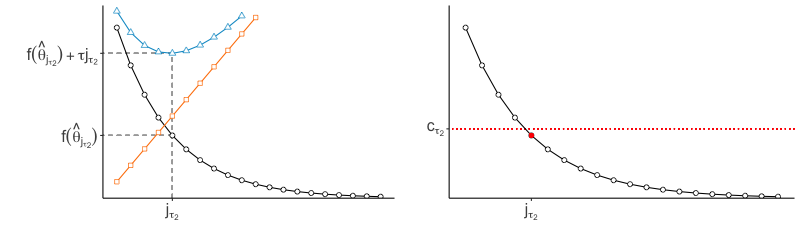}
}
\caption{An illustration of the scree plot approach for selecting the number of factors using (a) a small rank penalty $\tau_1$ and (b) a large rank penalty $\tau_2$. Black circles represent the non-increasing function $j \mapsto f(\hat{\theta_j})$, orange squares represent the rank penalty $\tau j$, and blue triangles represent their sum (the function to be minimized). The number $c_{\tau}$ is one example of a threshold furnishing the rank $j_{\tau}$.}
\label{fig:scree-illust}
\end{figure}

We solve the optimization problems in (a) by formulating them as low-rank covariance matrix completion (LRCMC) problems. Note that any rank-$j$ covariance tensor $\theta \in \mbb{R}^{\mfrak{M}\times\mfrak{M}}$ has square matricization $\theta_{\text{sq}} \in \mbb{R}^{M \times M}$, $M = M_1 \times \dots \times M_D$, that may be expressed as $\mb{V}\mb{V}^T$, where $\mb{V} \in \mbb{R}^{M\times j}$. The optimization problems in (a) thus reduce to the factorized matrix completion problems
\begin{equation}
\label{eqn:est-lrcmc}
    \min_{\mb{V}\in \mbb{R}^{M\times j}} \norm{ \mb{Z}_{\text{sq}} \circ \left( \hat{\mcal{C}}^s_{\text{sq}} - \mb{V}\mb{V}^T \right) }_F^2,
\end{equation}
for $j = 1, \dots, K^*$. Although the problems in (\ref{eqn:est-lrcmc}) are not convex in $\mb{V}$, there is reason to trust that gradient-based algorithms will converge to ``good'' local minima. One reason is that closely related problems have no spurious local minima (i.e., all local minima are global; \citealp{ge-etal-2016}). Other theory suggests that gradient-based routines converge to statistically useful solutions given high-quality initializations \citep{chen-wainwright-2015}. Bearing this in mind, \cite{stanley-etal-2025} solved the problems in (\ref{eqn:est-lrcmc}) using an implementation of the Broyden–Fletcher–Goldfarb–Shanno (BFGS; \citealp{broyden-1970,fletcher-1970,goldfarb-1970,shanno-1970}) algorithm, which they initialized with $\mb{V}_0 = \mb{U}_j \mb{\Sigma}_j^{1/2}$, where $\mb{U}_j \mb{\Sigma}_j \mb{U}_j^T$ is the rank-$j$ truncated eigendecomposition of $\hat{\mcal{C}}^s_{\text{sq}}$. Although this approach indeed discovers ``good'' local minima, its application does not scale to the full brain. Consider that the 300 thousand voxels in each AOMIC resting-state scan yield a spatial covariance that consumes 365 GB when stored as 32-bit floating points. One can shrink this problem by considering only those voxels that reside within the brain but even then the spatial covariance at hand is massive. To start, this means that the eigendecompositon of $\hat{\mcal{C}}^s_{\text{sq}}$ required for initialization is prohibitively expensive. We work around this by applying a randomized SVD solver \citep{halko-etal-2011} to a subset of the original data that allows us to approximate the initialization of \cite{stanley-etal-2025}. However, the central challenge in solving (\ref{eqn:est-lrcmc}) remains: any single-machine algorithm that processes the full spatial covariance at once will suffer from slow iterations and, even worse, exceed the memory limits of many machines. We address this by invoking a distributed stochastic gradient descent (DSGD; \citealp{stanley-etal-2025}) algorithm that solves a wide class of LRCMC problems, including the one in (\ref{eqn:est-lrcmc}). This algorithm, which adapts the procedure of \cite{gemulla-etal-2011} to covariance matrices, requires little inter-machine communication, and distributes the storage of large covariance matrices across multiple machines.

\subsection{Subspace Exploration}
\label{sec:est-subexp}

After estimating the global subspace, the next task is to choose a representation of this subspace. Of course, we could always select the initial loadings $\hat{\mcal{L}}_k$, but these generally obscure interesting structure. This is because eigentensors tend to have many high-magnitude weights, which complicates their identification with parsimonious spatial maps that arise from neurological processes (e.g., visual processing). The problem is that eigentensors must be orthogonal, a restriction that often scrambles parsimonious structure hidden within the global subspace. Rotation can unscramble this structure, and produce interpretable loading estimates that greatly facilitate interpretation of the latent factors. Rotations may be orthogonal or oblique, the former preserving uncorrelated factors, and the latter inducing correlations among them. Below, we describe several tensorial rotation procedures whose implementation reduces to transforming the matricized tensors via a multivariate procedure. In these descriptions, let $\hat{\mb{L}} \in \mbb{R}^{M \times \hat{K}}$, $M=M_1, \dots, M_D$, be the matrix representation of the loading tensors whose $k$th column is the vectorization of the initial estimate of the $k$th loading tensor.

\subsubsection{Orthogonal Rotation}
\label{sec:est-subexp-or}

Orthogonal rotation is appropriate when one would like to study uncorrelated (on average) functional factors. There are numerous orthogonal rotation algorithms for multivariate data, each choosing some $\hat{K}$-by-$\hat{K}$ orthonormal matrix $\mb{R}$ that rotates the initial loadings towards some optimal structure. Two of the most popular orthogonal rotation procedures are quartimax \citep{neuhaus-wrigley-1954} and varimax \citep{kaiser-1958}, which simplify the rows and columns, respectively, of the loading matrix. If we define $\hat{\und{\mcal{L}}}^*(\und{m}) = \hat{\und{\mcal{L}}}(\und{m}) \mb{R}$, then in the tensorial setting, quartimax maximizes
\begin{equation}
\label{eqn:est-fr-quart}
    \sum_{k=1}^K  \norm{(\hat{\mcal{L}}_k^*)^4}_F ,
\end{equation}
while varimax maximizes
\begin{equation}
\label{eqn:est-fr-var}
    \sum_{k=1}^K  \norm{(\hat{\mcal{L}}_k^*)^4}_F - \norm{(\hat{\mcal{L}}_k^*)^2}_F^2 ,
\end{equation}
both with respect to $\mb{R}$. Practically, maximizing (\ref{eqn:est-fr-quart}) and (\ref{eqn:est-fr-var}) is equivalent to applying multivariate quartimax and varimax to $\hat{\mb{L}}$, respectively.

\subsubsection{Oblique Rotation}
\label{sec:est-subexp-ob}

If one wishes to discover a structure simpler than that found by orthogonal rotation and is not constrained to studying uncorrelated (on average) functional factors, then one should use oblique rotation. Each oblique rotation scheme identifies some $\hat{K}$-by-$\hat{K}$ invertible matrix $\mb{T}$ satisfying $\text{diag}(\mb{T}\mb{T}^T) = \mb{I}$ that rotates the loading matrix towards a structure amenable to interpretation. Among the most popular is direct oblimin rotation \citep{jennrich-sampson-1966} which encourages simplicity while controlling factor correlation. If we define $\hat{\mcal{L}}_k^* (\und{m}) = \hat{\mcal{L}}_k (\und{m}) \mb{T}^{-1}$, then in the tensorial setting, direct oblimin minimizes 
\begin{equation}
\label{eqn:est-fr-oblimin}
    \sum_{k < k'} \norm{(\hat{\mcal{L}}_k^*)^2 (\hat{\mcal{L}}_k^*)^2}_F - \alpha \norm{(\hat{\mcal{L}}_k^*)^2}_F \norm{(\hat{\mcal{L}}_k^*)^2}_F,
\end{equation}
as a function of $\mb{T}$, where $\alpha$ controls the degree to which factors are correlated. When $\alpha$ is zero, the factors are maximally correlated, and direct oblimin is called \textit{quartimin}. As $\alpha$ takes on increasingly negative values, the factors become less correlated. Generally, one should avoid using positive values of $\alpha$ as this may produce degenerate solutions. In practice, we minimize (\ref{eqn:est-fr-oblimin}) by applying multivariate direct oblimin rotation to $\hat{\mb{L}}$.

\subsection{Postprocessing}
\label{sec:est-pp}

After selecting a useful representation of the global subspace, we may (optionally) postprocess this representation via smoothing and shrinkage. These steps can refine the estimate of the global spatial covariance and lead to more interpretable representations.

\subsubsection{Smoothing}
\label{sec:est-pp-smooth}

Since the objective of (\ref{eqn:est-g-fr}) ignores contaminated on-band elements of the empirical covariance, its minimization implicitly smooths the initial $\hat{\mcal{L}}_k$. In practice, however, these estimates may still be too rough, and could benefit from explicit smoothing. \cite{stanley-etal-2025} encourage such smoothness by adding a roughness penalty to their objective function. We instead recommend smoothing the rotated $\hat{\mcal{L}}_k^*$ as a postprocessing step. Though less streamlined than the strategy of \cite{stanley-etal-2025}, postprocessed smoothing allows users to more easily achieve different levels of smoothness for each loading. We opt to smooth each loading independently with a Gaussian filter that applies one-dimensional convolutions along each spatial dimension. Component-wise smoothness is controlled by scale parameters $\sigma_k$ that set the standard deviation of the Gaussian kernels used in each one-dimensional convolution. Section \ref{sec:supp-tune} of the supplement describes how to tune these $\sigma_k$ via cross-validation. Henceforth, we let $\Tilde{\mcal{L}}_k$ denote the $k$th rotated loading after smoothing.

\subsubsection{Shrinkage}
\label{sec:est-pp-shrink}

Rotation induces a ``simple'' loading structure which, roughly speaking, is achieved when there are many low-magnitude weights and relatively few high-magnitude weights. However, some applications call for a ``sparse'' loading structure in which many weights are not just low-magnitude, but are actually set to zero. This is arguably the case in FC wherein it is useful to know which brain regions are not activated by a latent factor. Our shrinkage procedure attains sparsity by adaptively soft-thresholding the smooth rotated $\Tilde{\mcal{L}}_k$: 
\begin{equation*}
    \bar{\mcal{L}}_k (\und{m}) = \text{sgn}\left(\Tilde{\mcal{L}}_k (\und{m})\right) \max \left( \abs{\Tilde{\mcal{L}}_k (\und{m})} - \kappa_k w\left(\Tilde{\mcal{L}}_k (\und{m})\right) , 0\right),
\end{equation*}
where $w(\cdot)$ is a weight function with positive support (e.g., $w(x) = \abs{x}^{-2}$), and the $\kappa_k > 0$ are shrinkage parameters which may be tuned via cross-validation as described in Section \ref{sec:supp-tune} of the supplement.

\subsection{Factor Score Estimation}
\label{sec:est-fse}

Though the discretized factors $\mcal{F}_1, \dots, \mcal{F}_{\hat{K}} \in \mbb{R}^J$ are random, and not parameters in the usual sense, their estimates, called \textit{factor scores}, can be quite informative. In fMRI, these estimates track the subject-specific activity of estimated spatial maps throughout a scan. They can also serve as inputs to diagnostic tests that evaluate the suitability of the assumption in (\ref{eqn:ir-assump-facs-or}), that the factors are uncorrelated with unit variance on average. 

We use techniques from functional regression to compute factor scores. Let us temporarily return to the infinite-resolution TFFM, but fix the spatial location at one of the discrete spatial observation points $\und{s}_{\und{m}}$ and swap out the true loadings $l_k$ evaluated at that point for the postprocessed estimates $\bar{\mcal{L}}_k$. The resulting model may be written as
\begin{equation*}
    X_{\und{m}}(t) = \sum_{k=1}^{\hat{K}} \bar{\mcal{L}}_{k,\und{m}} F_k(t) + \epsilon_{\und{m}}(t),
\end{equation*}
where $X_{\und{m}}(t) = X(\und{s}_{\und{m}}, t)$, $\bar{\mcal{L}}_{k,\und{m}} = \bar{\mcal{L}}_{k}(\und{m})$, and $\epsilon_{\und{m}}(t) = \epsilon(\und{s}_{\und{m}}, t)$. That is, the infinite-resolution TFFM becomes a function-on-scalar (F-on-S) regression model where $X_{\und{m}}$ is the functional response indexed discretely over space, the $\bar{\mcal{L}}_{k,\und{m}}$ are viewed as scalar predictors, and the $F_k$ are the functional parameters to be estimated. In Section \ref{sec:supp-fse-est} of the supplement, we describe a F-on-S method for estimating these $F_k$ that is similar to the one outlined in Section 13.4.5 of \cite{ramsay-silverman-2005}. We control the smoothness of the estimates via a component-wise tuning parameter $\und{\gamma}$ that may be selected using the spatial cross-validation procedure described in Section \ref{sec:supp-fse-tune} of the supplement. With these functional estimates at hand, one need only evaluate each $\hat{F}_k$ on $\mcal{T}$ to obtain estimates for the discrete $\mcal{F}_k$.

\subsubsection{A Diagnostic for the Orthogonal TFFM}
\label{sec:est-fse-dt}

One may use factor score estimates $\hat{\mcal{F}}_i \in \mbb{R}^{\hat{K}\times J}$, $i = 1, \dots, n$, to validate the assumption in (\ref{eqn:ir-assump-facs-or}) of the orthogonal model, that the factors are, on average, uncorrelated with unit variance. If this assumption holds, then one would expect 
\begin{equation*}
    \frac{1}{nJ} \sum_{i=1}^n \hat{\mcal{F}}_i \hat{\mcal{F}}_i^T \approx \mb{I}_{\hat{K}\times \hat{K}}.
\end{equation*}
In Sections \ref{sec:aomic}, we employ this diagnostic to assess the suitability of the orthogonal TFFM when applied to resting-state fMRI data.

\section{Simulation Studies}
\label{sec:sim}

We now investigate the efficacy of our methodology through three simulation studies. In the first, we compare the accuracy of our postprocessed estimator for the global spatial covariance to that of several competing estimators. In the second, we highlight the different ways in which select methods express the estimated global subspace via the empirical loadings. In the third, we study the performance of our methodology for factor score estimation in various settings. We begin, in Section \ref{sec:sim-data}, by outlining the data generating process used across these studies. 

\subsection{Data Simulation}
\label{sec:sim-data}

For each study, we simulate data from a TFFM with spatial dimension equal to two, and simplify designs by setting $M = M_1 = M_2$ and $\delta = \delta_1 = \delta_2$. We simulate $n$ i.i.d. mean-zero functions $Y_i(\und{s},t)$ and $n$ i.i.d. mean-zero functions $\epsilon_i(\und{s}, t)$ on an evenly spaced $M \times M \times J$ lattice in $[0, 1]^3$, and summarize these discretizations with the order-3 tensors $\mcal{Y}^{(i)}$ and $\mcal{\epsilon}^{(i)}$, respectively. We sum these components to get the samples $\mcal{X}^{(i)} = \mcal{Y}^{(i)} + \mcal{\epsilon}^{(i)}$, then calculate the empirical spatial covariance tensor
\begin{equation*}
    \hat{\mcal{C}}_n^s = \frac{1}{nJ} \sum_{i=1}^n \sum_{j=1}^J \mcal{X}^{(i)}_{j} \otimes \mcal{X}^{(i)}_{j},
\end{equation*}
where $\mcal{X}^{(i)}_{j} \in \mbb{R}^{M \times M}$ is the 2-dimensional spatial image at the $j$th time point.

\begin{figure}[!h]
\centering
\subfloat[BI]{\label{fig:loads-bi}\includegraphics[width=0.35\linewidth, trim=0 40 0 40, clip]{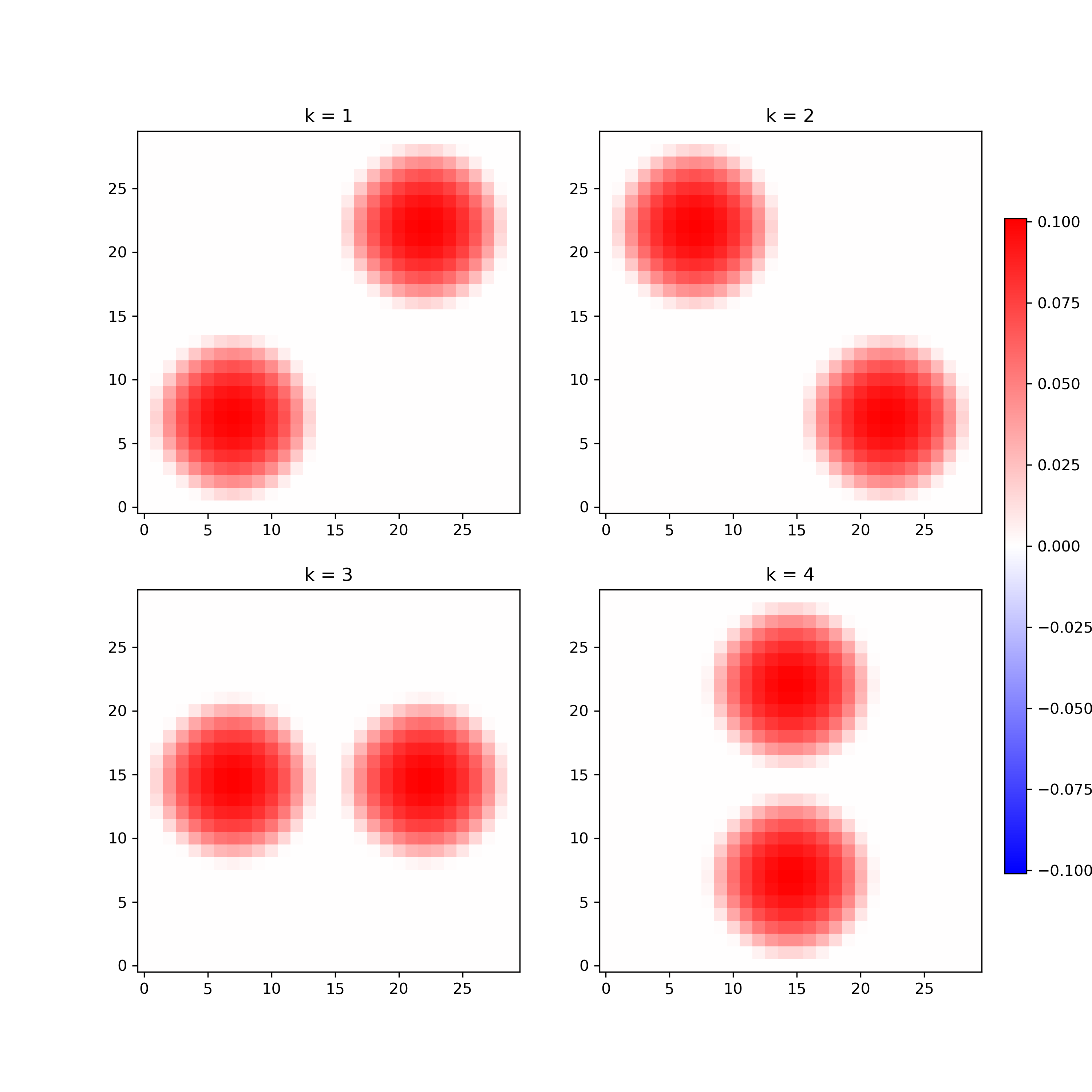}}\qquad
\subfloat[NET]{\label{fig:loads-net}\includegraphics[width=0.35\linewidth, trim=0 40 0 40, clip]{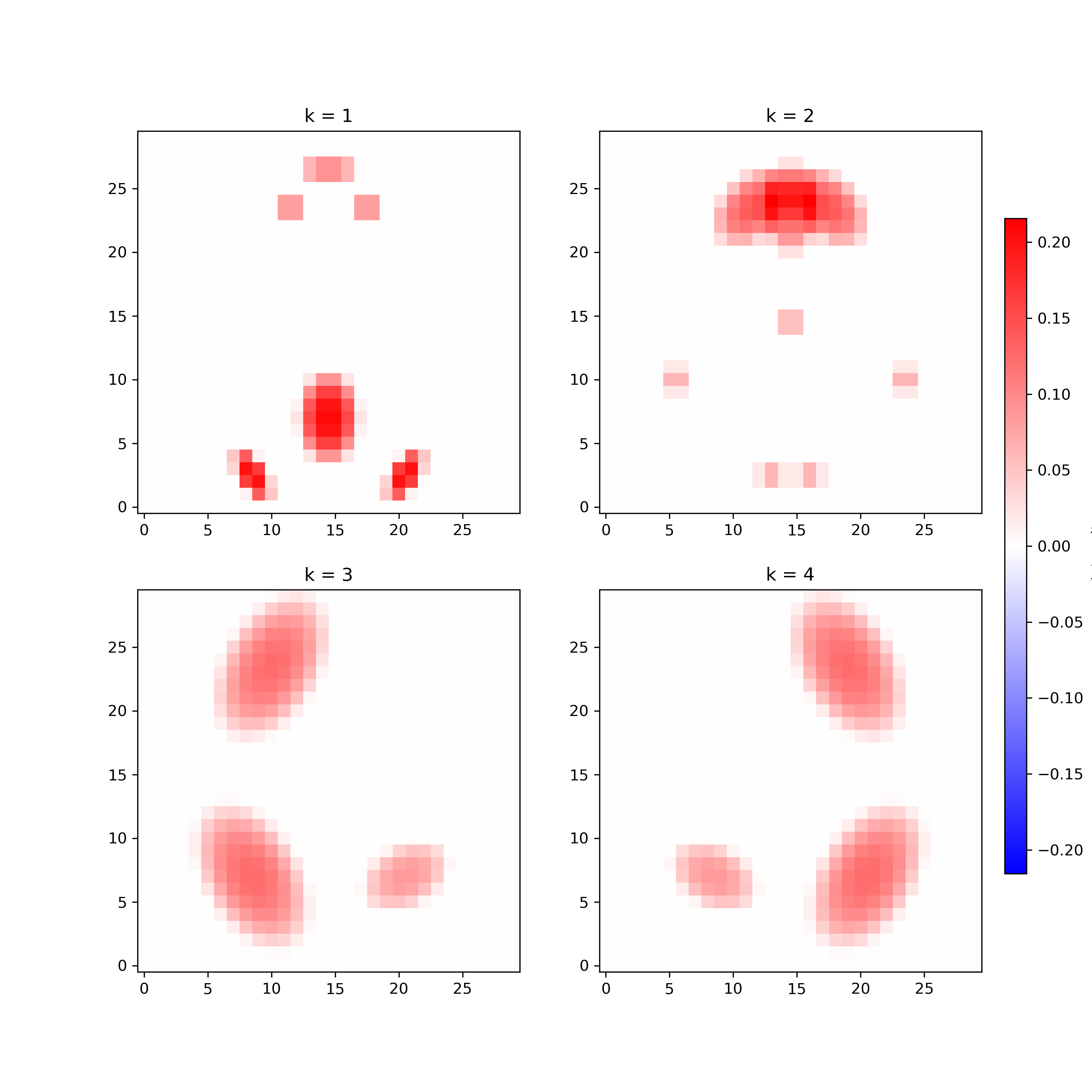}}\\
\subfloat[TRI]{\label{fig:loads-tri}\includegraphics[width=0.7\linewidth, trim=0 40 0 40, clip]{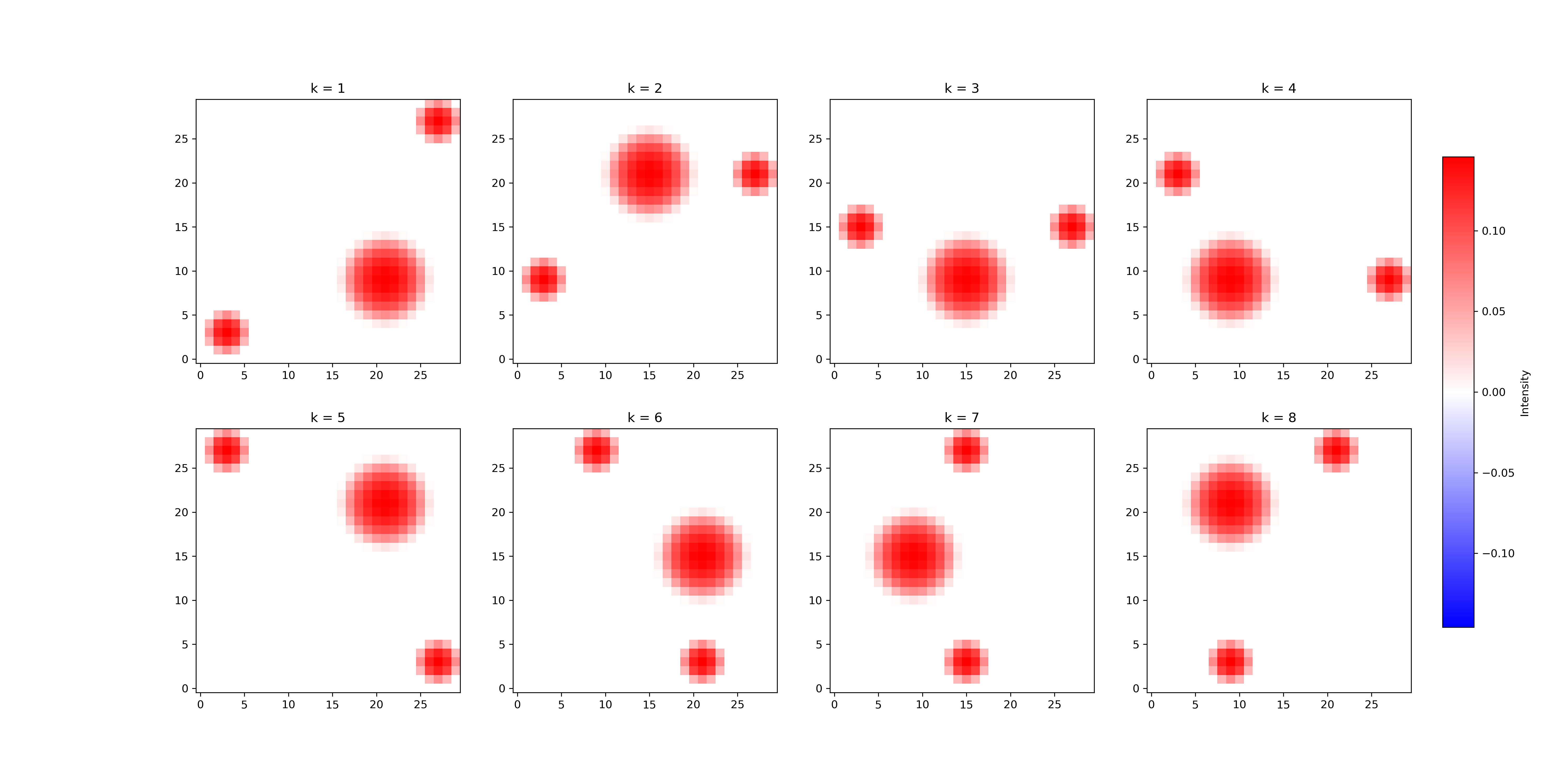}}
\caption{The three loading schemes used throughout simulation studies.}
\label{fig:load-schemes}
\end{figure}

To simulate the functions $Y_i$, we set $Y_i(\und{s}, t) = \sum_{k=1}^K c_k z_k(\und{s}) f_{ik}(t)$, where the $z_k$ are (possibly non-orthogonal) functions scaled to unit norm, the $c_k$ are positive constants, and the $f_{ik}$ are (in the orthogonal case) i.i.d. realizations of a zero-mean Gaussian process with squared exponential kernel having length parameter $\omega_f$ (n.b., such factors will have pointwise unit variance). In the oblique case, we rotate $\und{f}_i = [f_{i,1}, \dots f_{i,K}]$ by some invertible matrix $\pmb{T}$ satisfying $\text{diag}(\mb{T}\mb{T}^T) = \mb{I}$ to induce the covariance $\mb{H} = \mb{T}\mb{T}^T$ among the factors. The $k$th loading function is decomposed into its normalization $z_k$ and a scaling constant $c_k$ to make some loadings functions more difficult to estimate than others. Moreover, the $(c_k, z_k)$ are not necessarily equal to the eigen-pairs $(\lambda_k, \eta_k)$ of the global spatial covariance $\mscr{G}$, as we do not require that the $z_k$ are orthogonal. Throughout these studies, we consider the three \textit{loading schemes} depicted in Figure \ref{fig:load-schemes}. In the first scheme (denoted by BI), each $z_k$ is defined piecewise with a bump function in two regions of its domain, and is zero elsewhere. Each $z_k$ in the second scheme (denoted by NET) is also defined by piecewise bump functions, but their arrangement mimics that of known resting-state networks (e.g., default mode, executive, right dorsal visual stream, left dorsal visual stream). In the third loading scheme (denoted by TRI), each $z_k$ is defined by three bump functions, one of which is larger than the others.

To simulate the functions $\epsilon_i$, we set $\epsilon_i (\und{s}, t) = \sum_{j=1}^P a_{ij} v_{ij}(\und{s}) u_{ij}(t)$, where the $u_{ij}$ are i.i.d. realizations of a zero-mean Gaussian process with squared exponential kernel having length parameter $\omega_{u}$, the $a_{ij}$ are positive constants, and $P >> K$. Furthermore, the (possibly non-orthogonal) $v_{ij}$ are 2-dimensional bump functions scaled to unit norm with support contained by $[(p-1)\delta, p\delta] \times [(q-1)\delta, q\delta]$ where $p$ and $q$ are chosen so that the intersection of this product with $[0,1]^2$ is non-empty. Note that the local spatial covariance may be different for each subject.

In all simulations, we set the number of time points $J = 500$, a number close to that of the AOMIC resting-state data (where, after pre-processing, $J = 470$). Moreover, we set the length parameters $\omega_f = 0.02$ and $\omega_u = 0.002$ as this results in relative global-local temporal smoothness that roughly mimics that of resting-state cognitive processes and low-frequency noise. Finally, the way in which we choose the loading coefficients $c_k$ and error coefficients $a_{ij}$ determines the ``regime'' of a simulation and impacts the difficulty of estimation. Throughout these studies, we make use of two regimes, the second presenting a more challenging estimation problem than the first. In both regimes, the $a_{ij}$ are distributed uniformly in $[0.1,1]$, while the $c_k$ are distributed uniformly in $[2,3]$ and $[0.8, 1.8]$ in regimes one and two, respectively.

\subsection{Study 1: Global Covariance Estimation}
\label{sec:sim-s1}

The goal of our first study is to compare the accuracy of our postprocessed estimator (denoted by TFFA) to that of several other global covariance estimators: 
\begin{enumerate}
    \item TFFA without shrinkage (denoted by MCS, for marix completion with smoothing).
    \item TFFA with neither smoothing nor shrinkage (denoted by MC, for matrix completion).
    \item Independent component analysis with smoothing (denoted by ICAS): after pre-smoothing the data using a Gaussian filter, we estimate $\mcal{G}^s$ by extracting from the data an $M\times M\times K$ tensor $\hat{\mcal{S}}$ of independent components via MELODIC, then compute $\sum_k \hat{\mcal{S}_k} \otimes \hat{\mcal{S}_k}$ where the $M\times M$ matrix $\hat{\mcal{S}_k}$ is the $k$th estimated independent component.
    \item ICAS without pre-smoothing (denoted by ICA).
\end{enumerate}

For TFFA, MCS and ICAS, we must select smoothing and (possibly) shrinkage parameters prior to model estimation. To accomplish this, we utilize the cross-validation procedures outlined in Section \ref{sec:supp-tune} of the supplement, setting the number of folds equal to three. We expedite tuning by replacing component-wise hyperparameters (e.g., $\und{\sigma}$ and $\und{\kappa}$) with single parameters applied to each component (e.g., $\sigma$ and $\kappa$). When estimating via TFFA, MCS, or MC, we set the bandwidth equal to $0.1$, even when the true bandwidth is smaller. 

\begin{figure}[!h]
    \centering
    \includegraphics[scale=0.5]{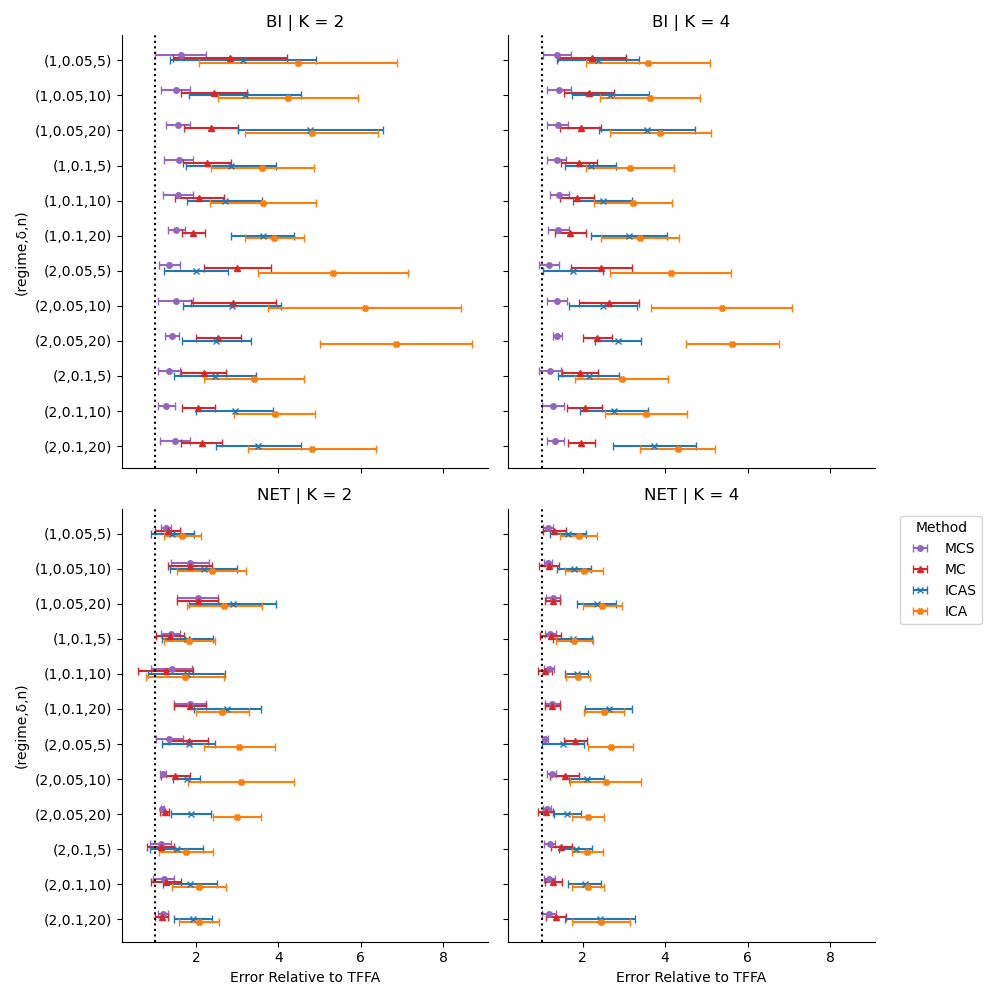}
    \caption{Mean relative errors (with $\pm 2$-standard-deviation error bars) for all scenarios using the TFFA estimator as baseline.}
    \label{fig:sim-est}
\end{figure}

The study evaluates each estimator in 48 scenarios, each defined by a loading scheme (BI or NET), number of factors (2 or 4), bandwidth (0.05 or 0.1), regime (1 or 2), and number of subjects (5, 10, or 20). For 100 repetitions of each scenario, we use each of the five estimators to compute an estimate $x$ of the global spatial covariance $\mcal{G}^s$ then compute normalized errors $E(x) = \norm{\mcal{G}^s - x}_F / \norm{\mcal{G}^s}_F$. To quantify the relative performance of the TFFA estimator, we form relative errors by dividing the normalized errors for the MCS, MC, ICAS, and ICA estimators by that for the TFFA estimator. Figure \ref{fig:sim-est} displays results for all scenarios. 

In nearly every scenario, TFFA furnishes the most accurate global covariance estimates. The benefits of shrinkage and smoothing (quantified by the gaps between the dotted line and MCS errors, and MCS and MC errors, respectively) are consistently present in BI scenarios but more sporadic in NET scenarios. This is likely because the universal smoothing parameters used in this study are poorly suited to the NET scheme whose loadings display variable sparsity and smoothness. This suggests that practitioners should opt for component-wise tuning when compute time is not a concern. 

The ICA-based estimators generally lag behind the others, although pre-smoothing (quantified by the difference between ICAS and ICA errors) is sometimes quite helpful, particularly in the second regime. In this regime, the benefits of pre-smoothing are most pronounced when $\delta = 0.05$ but less so when $\delta = 0.1$. This is because smoothing the data smears more unwanted local variation into the global covariance in high-bandwidth scenarios.

\subsection{Study 2: Global Covariance Expression}
\label{sec:sim-s2}

In our second study, we examine how various methods express their target subspace and comment on the alignment of these expressions with the ground truth. To do so, we examine output from orthogonal TFFA, oblique TFFA, and ICAS when data arise from orthogonal and oblique TFFMs, and model dimension is correctly specified (i.e., $\hat{K} = K$). In Section \ref{sec:supp-sim-s2} of the supplement, we explore the implications of dimension overspecification (i.e., $\hat{K} > K$). Throughout, we consider regime-two scenarios with loading scheme set to TRI, $K$ set to 8, $n$ set to 20, and $\delta$ set to $0.1$.

\begin{figure}[!h]
\centering
\subfloat[Orthogonal TFFA]{\label{fig:sim-exp-8-or_ffa-or}\includegraphics[width=0.6\linewidth]{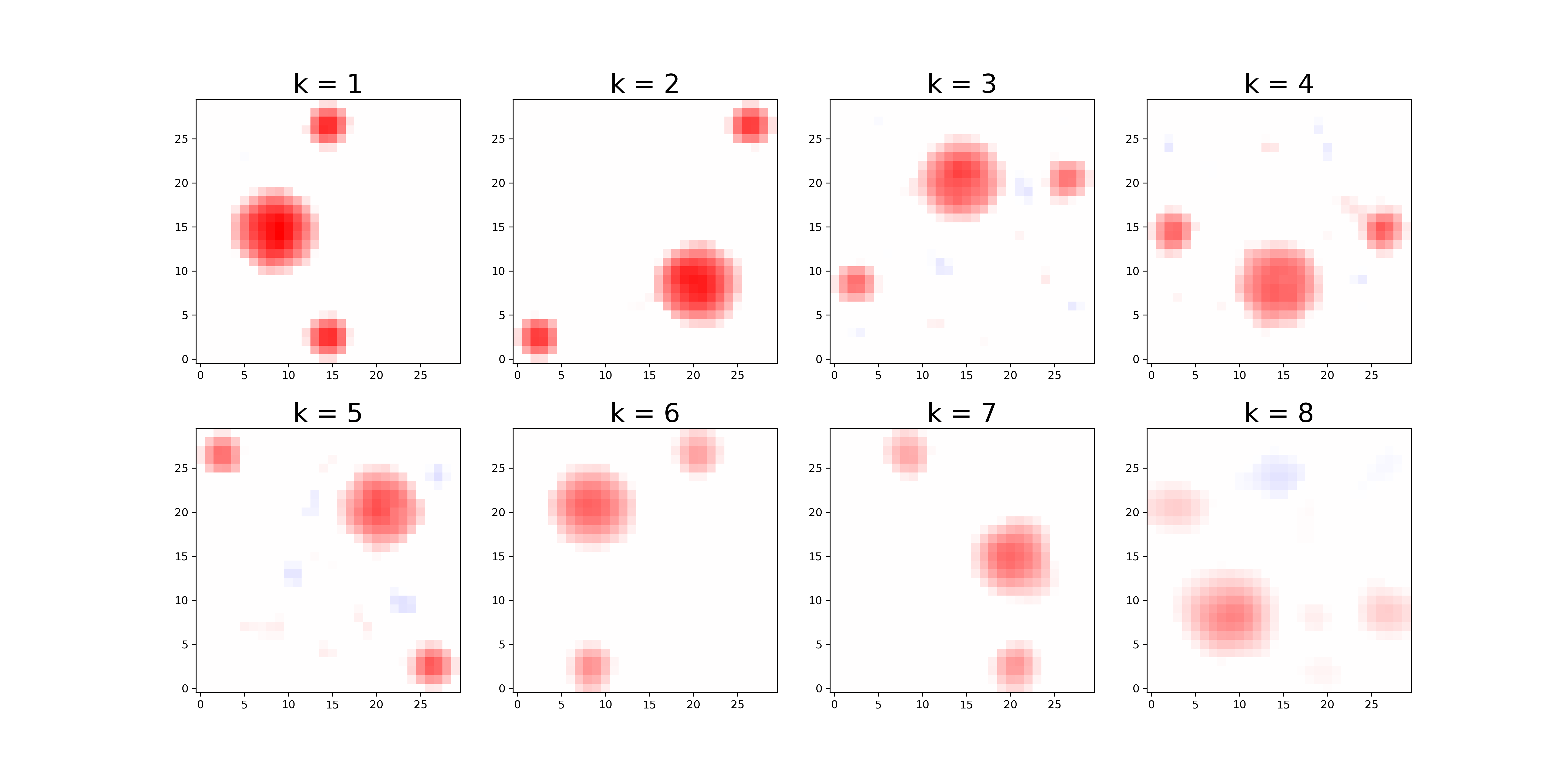}}\\
\subfloat[Oblique TFFA]{\label{fig:sim-exp-8-or_ffa-ob}\includegraphics[width=0.6\linewidth]{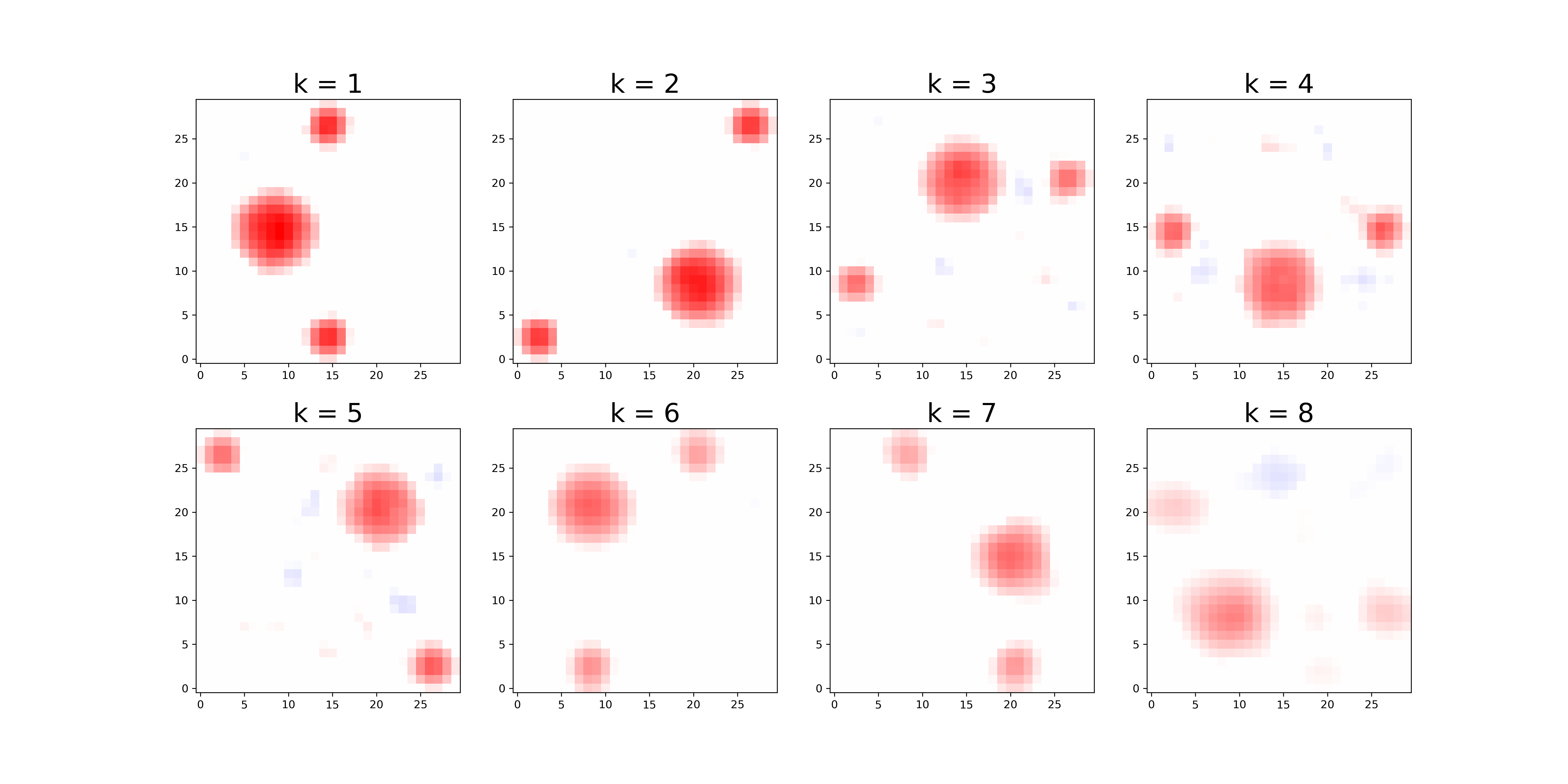}}\\
\subfloat[ICAS]{\label{fig:sim-exp-8-or_icas}\includegraphics[width=0.6\linewidth]{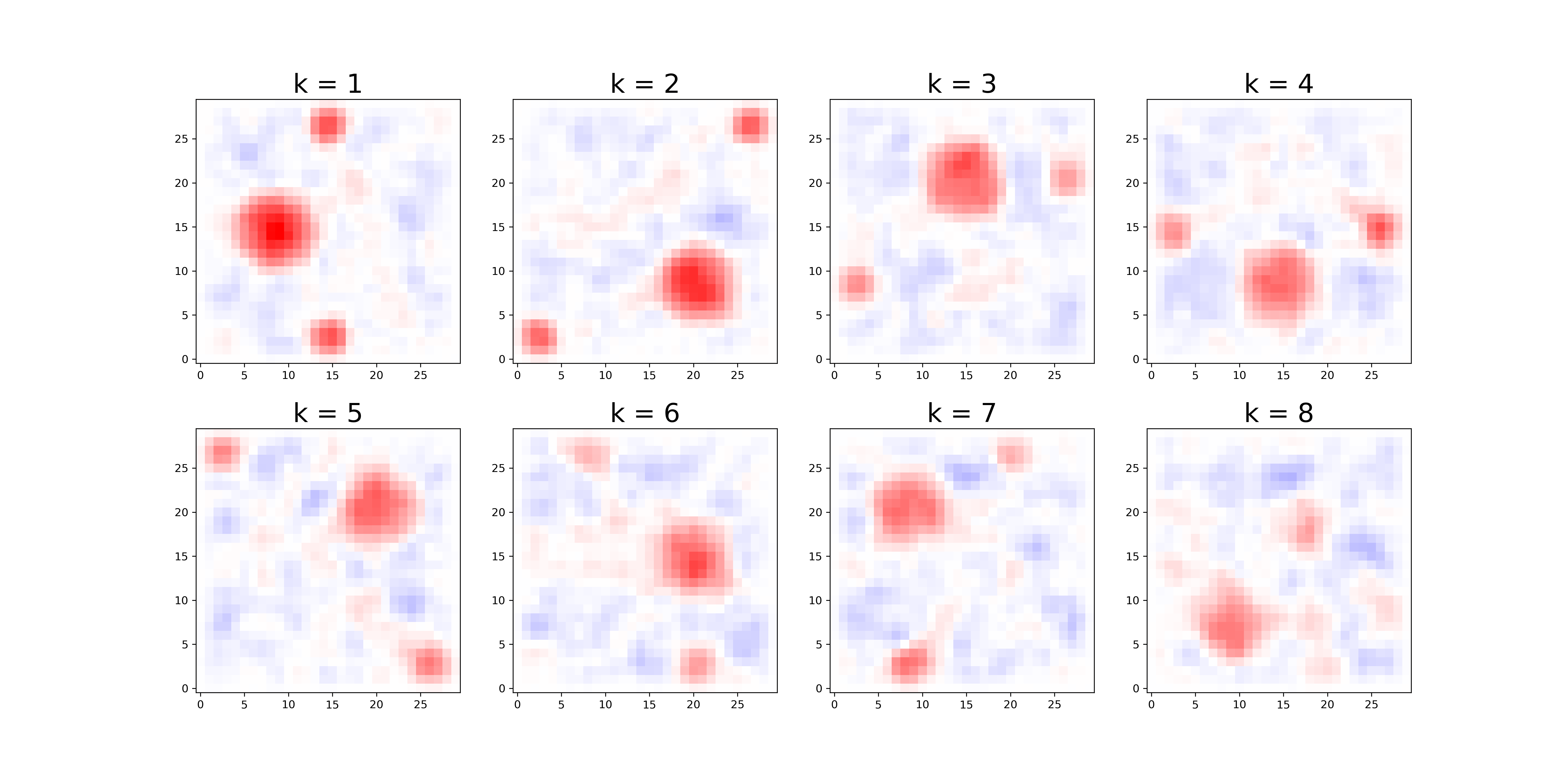}}
\caption{Estimates for order-8 models when data are generated from an orthogonal model in scenario of Study 2.}
\label{fig:sim-exp-8-or}
\end{figure}

\begin{figure}[!h]
\centering
\subfloat[Orthogonal TFFA]{\label{fig:sim-exp-8-ob_ffa-or}\includegraphics[width=0.6\linewidth]{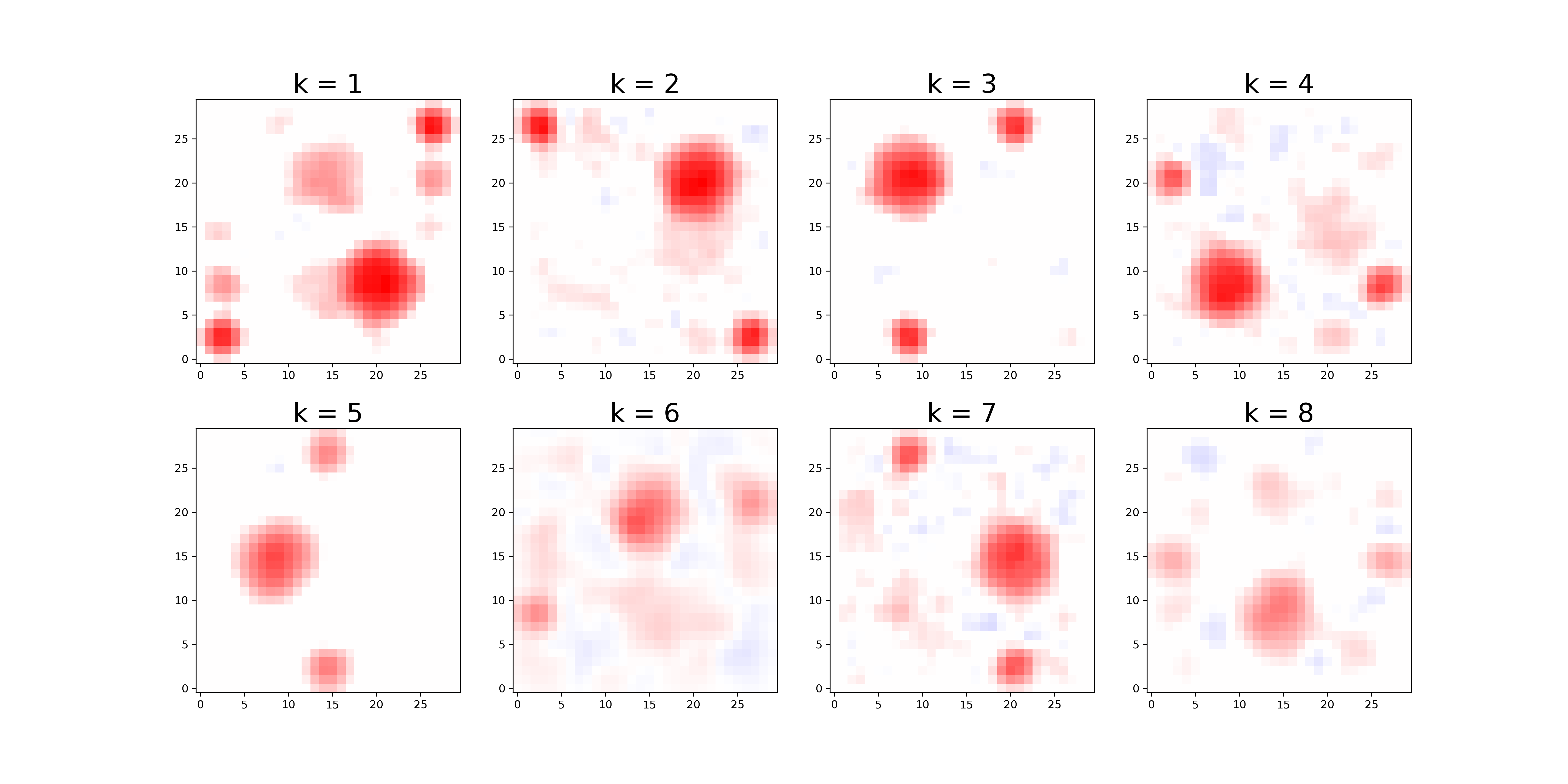}}\\
\subfloat[Oblique TFFA]{\label{fig:sim-exp-8-ob_ffa-ob}\includegraphics[width=0.6\linewidth]{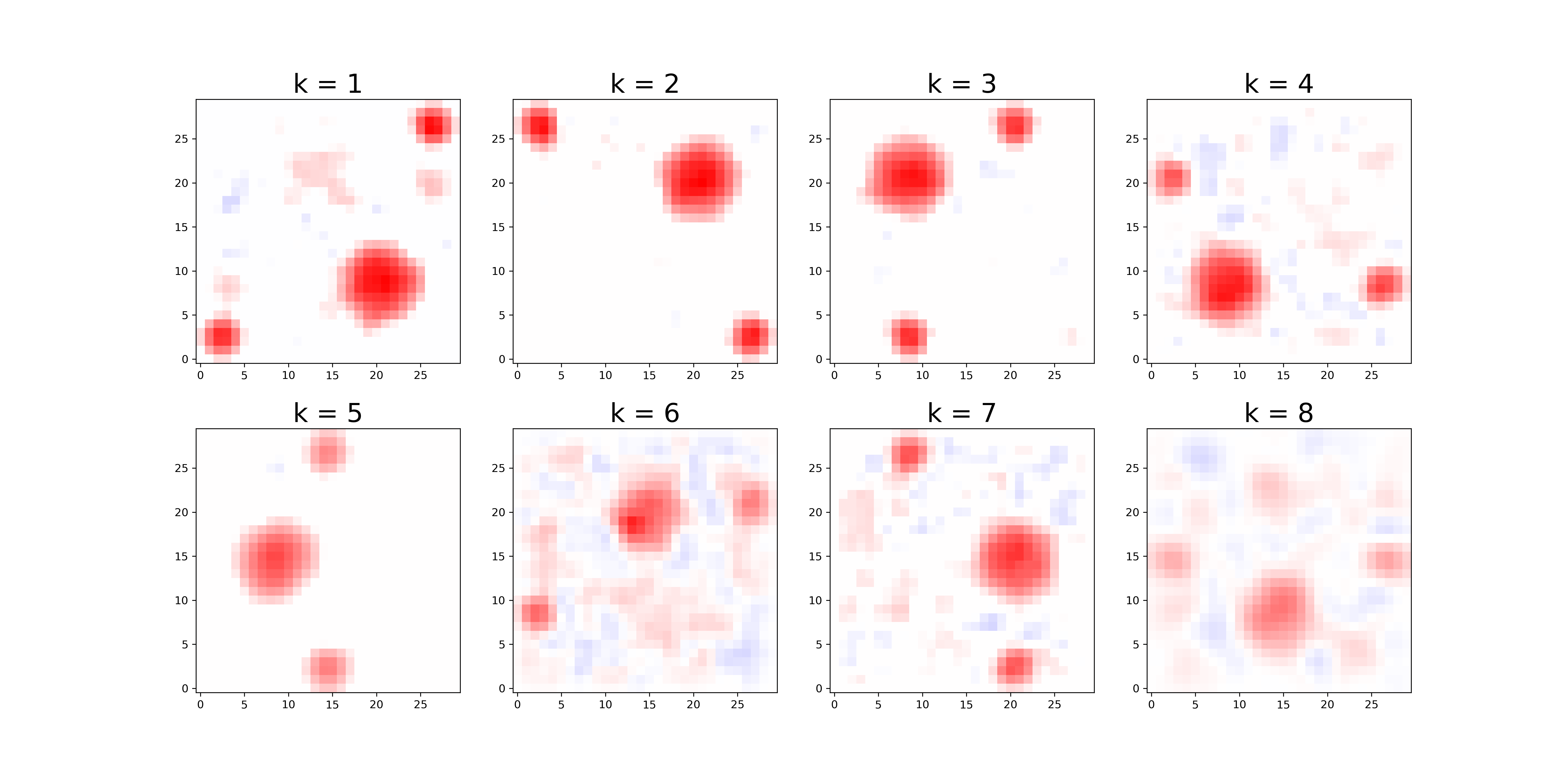}}\\
\subfloat[ICAS]{\label{fig:sim-exp-8-ob_icas}\includegraphics[width=0.6\linewidth]{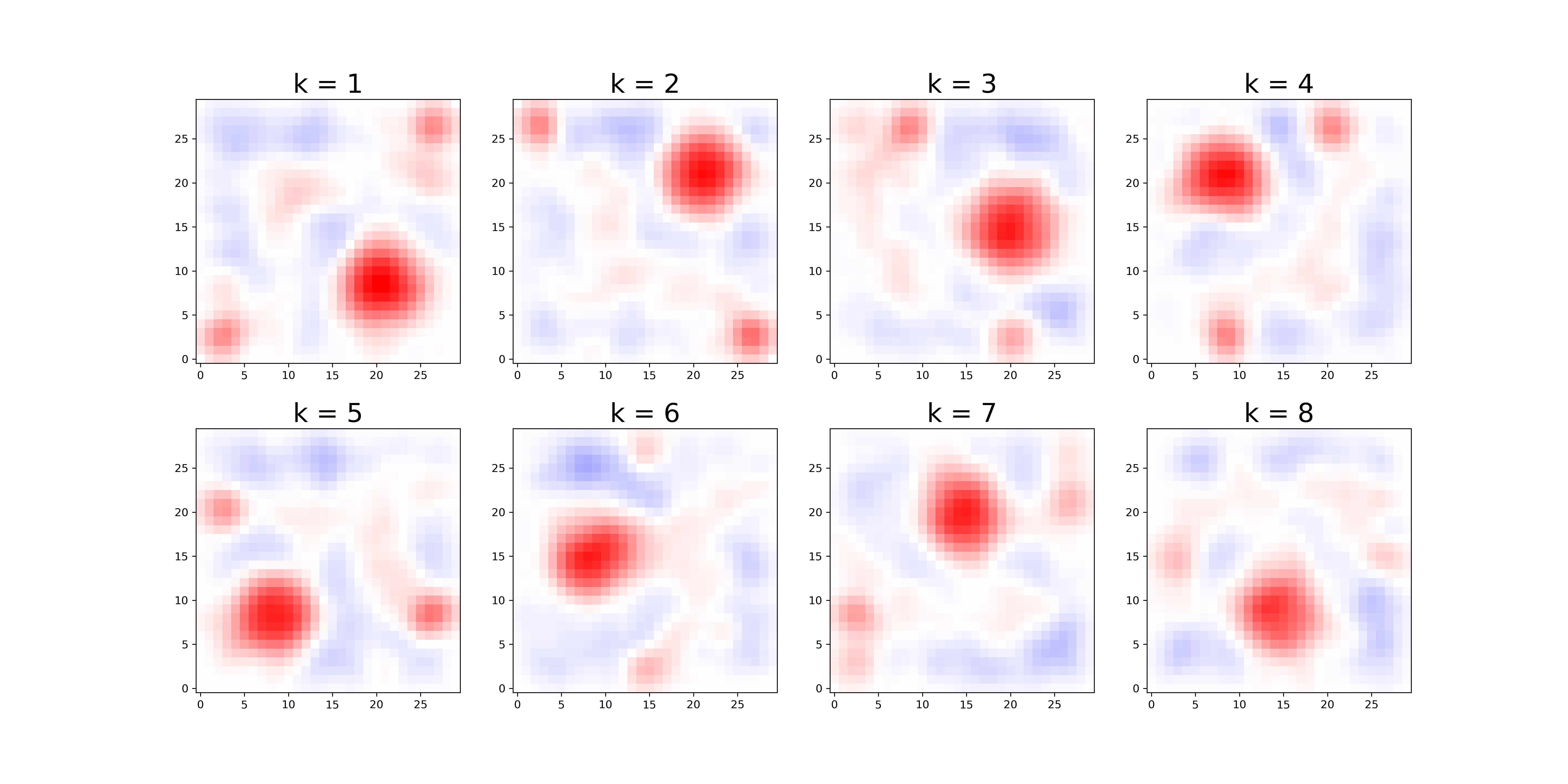}}
\caption{Estimates for order-8 models when data are generated from an oblique model in scenario of Study 2.}
\label{fig:sim-exp-8-ob}
\end{figure}

When data arise from an orthogonal model, estimates from all three methods are in close alignment with the true loading functions, while those from both TFFA methods possess sparse structure that facilitates interpretation (see Figure \ref{fig:sim-exp-8-or}). It is worth noting that despite employing an oblique rotation procedure, oblique TFFA is nonetheless able to recover the loadings of an orthogonal model since every orthogonal rotation is also a permissible oblique rotation. When data are generated from an oblique model, output from TFFA and ICAS is in reasonable alignment with the ground truth, but that from orthogonal TFFA is noticeably less simple than the true loadings (see Figure \ref{fig:sim-exp-8-ob}). In fact, since the space of orthogonal rotations is a strict (and comparatively tiny) subset of oblique rotations, oblique algorithms will, in general, discover simpler structure than orthogonal algorithms. It is thus advisable, at least when $K$ is known, to favor oblique TFFA over orthogonal TFFA unless one wishes to study only uncorrelated factors.

\subsection{Study 3: Factor Score Estimation}
\label{sec:sim-s3}

The third and final study evaluates the performance of our methodology for factor score estimation. Recall that to compute a subject's factor scores, we treat the postprocessed loadings as fixed predictors, view the subject's factors as functional parameters, then estimate these factors via function-on-scalar (F-on-S) regression. To ensure that error incurred during loading estimation does not contaminate this study, we replace the postprocessed loadings in the usual procedure with the true ones. 

Throughout this section, we consider regime-one data simulated from an orthogonal TFFM on a 40-by-40 grid using the BI loading scheme with $K=2$ and $n=20$. We evaluate factor score estimation in six scenarios, each given by the height $h$ (20, 30, or 40 pixels) of a  square mask centered within the 40-by-40 grid outside of which pixels are ignored, and a bandwidth $\delta$ (0.05 or 0.1).

\begin{figure}[!h]
    \centering
    \includegraphics[width=0.5\linewidth]{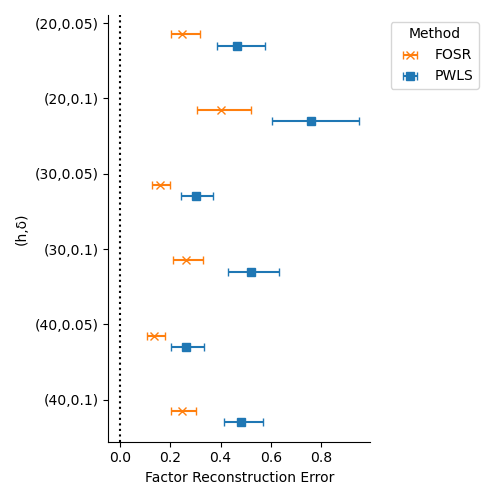}
    \caption{Factor reconstruction errors for the FOSR and PWLS estimators for each combination of $h$ and $\delta$.}
    \label{fig:sim-fse}
\end{figure}

To isolate the effect of smoothing, we compare our approach (denoted by FOSR, for F-on-S regression) to one that estimates the factors at each time point independently (denoted by PWLS, for pointwise least squares). For 10 repetitions of each $(h,\delta)$ combination, we estimate factor scores via FOSR and PWLS. We then compute $nK$ reconstruction errors for each repetition by evaluating $E_{ik} (x) = \norm{\und{F}_{ik} - x}_F \ \norm{\und{F}_{ik}}_F$ at the $i$th subject's $k$th factor score estimate $x$, for $i = 1, \dots, n$ and $k = 1, \dots K$. Figure \ref{fig:sim-fse} plots these errors for each scenario and method.

In all scenarios, FOSR errors are substantially smaller than PWLS errors, providing clear evidence for the benefits of smoothing. Moreover, the estimates for both methods improve as $h$ increases. In fMRI, increasing $h$ is analogous to analyzing a larger region of the brain (i.e., expanding a study's brain mask). Note that this is not equivalent to increasing the spatial resolution of a fixed region, which only slightly increases the effective sample size and thus has little effect on the quality of factor score estimates. However, estimates for both methods degrade as $\delta$ increases. This is because neither method accounts for spatial autocorrelation among the residuals. One could address this by instead minimizing some generalized least squares criterion, but this would necessitate the inversion of a large spatial error covariance, an intractable task in our fMRI application.

\section{AOMIC Analysis}
\label{sec:aomic}

We now turn to the AOMIC-PIOP1 dataset which includes six-minute resting-state scans for 210 healthy university students. Before each subject's resting-state session, a structural image was acquired at a resolution of 1 mm isotropic. A functional image was then acquired at a spatial resolution of 3 mm isotropic and a temporal resolution of 0.75 seconds. Preprocessing of anatomical and functional MRI were performed using a standard Fmriprep pipeline \citep{esteban-etal-2019}, as detailed in \cite{snoek-et-al-2021}.

We conducted three additional preprocessing steps to prepare these data for analysis via our methodology. To start, we discarded the first ten volumes from each subject's scan as these often contained large signal fluctuations due to scanner instability. Next, we used FMRIB's ICA-based X-noiseifier (FIX; \citealp{salimi-etal-2014, griffanti-etal-2014}) to remove lingering noise components (e.g., head movement, respiratory motion, and scanner artifacts) from each subject's scan that remained after Fmriprep processing. This involved (i) performing single-subject ICA via MELODIC for all 210 subjects, (ii) manually labeling noise components for 10 subjects, (iii) training the FIX classifier on this hand-labeled data, (iv) automatically identifying noise components with this classifier for the remaining 200 subjects, then (v) regressing out noise components from each subject's scan. When hand-labeling ICA components, we adhered to well-established criteria described in \cite{griffanti-etal-2014}. FIX denoising is now a workhorse technique in the neuroimaging literature, and has been incorporated into several major data preparation pipelines, like those of the Human Connectome Project \citep{glasser-2016}. Finally, to prevent high-variation voxels from dominating analysis, we applied variance normalization to all voxels time courses. 

\begin{figure}[!h]
\centering
\subfloat[Orthogonal TFFA]{\label{fig:aomic-210_ffa-or}\includegraphics[width=0.35\linewidth]{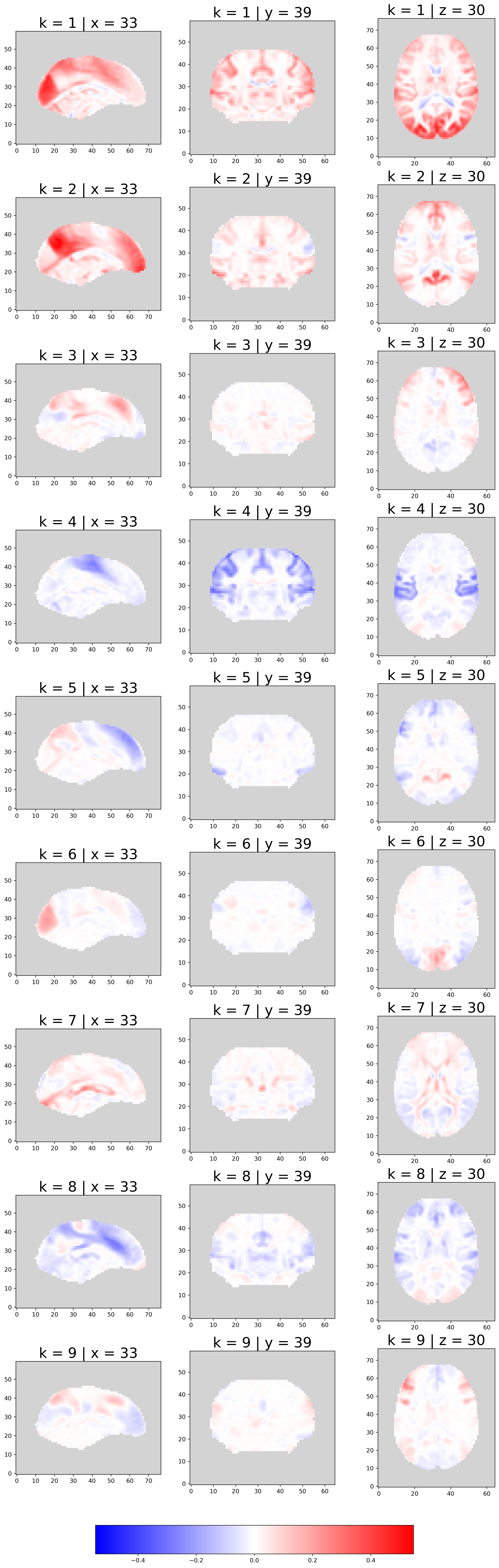}}\qquad
\subfloat[Oblique TFFA]{\label{fig:aomic-210_ffa-ob}\includegraphics[width=0.35\linewidth]{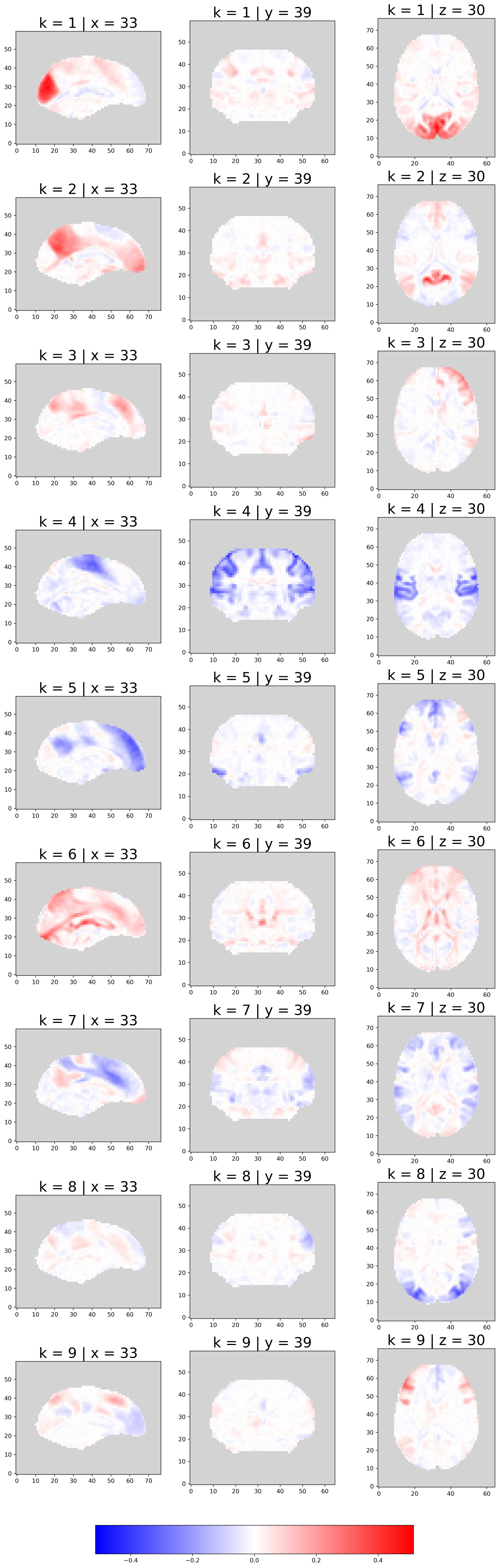}}
\caption{Loading estimates for the orthogonal and oblique TFFMs fit to data from 210 AOMIC subjects.}
\label{fig:aomic-210_ffa}
\end{figure}

\begin{figure}[!h]
\centering
\subfloat[Orthogonal TFFA]{\label{fig:aomic-210_fac-cov_ffa-or}\includegraphics[width=0.45\linewidth]{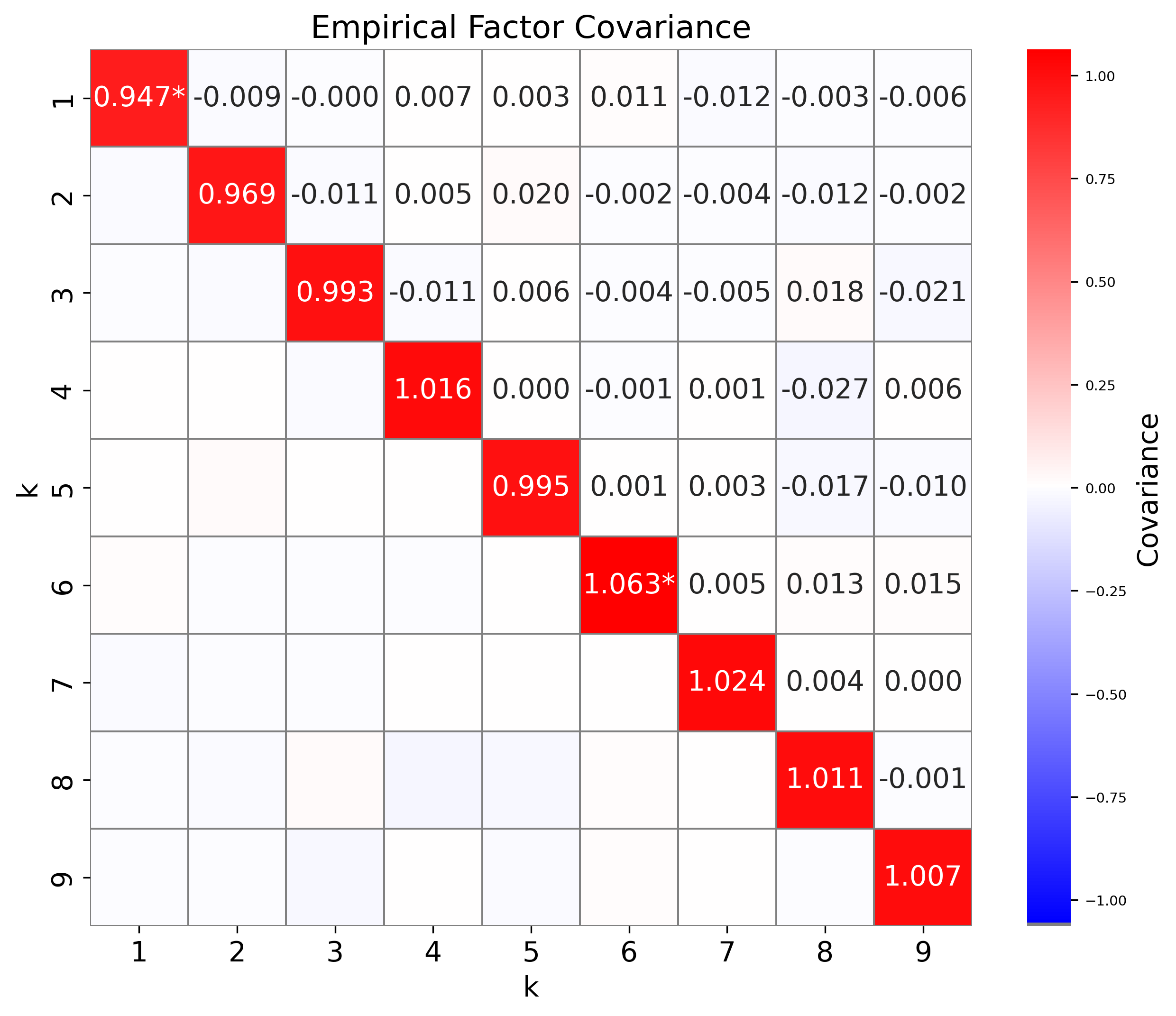}}\qquad
\subfloat[Oblique TFFA]{\label{fig:aomic-210_fac-cov_ffa-ob}\includegraphics[width=0.45\linewidth]{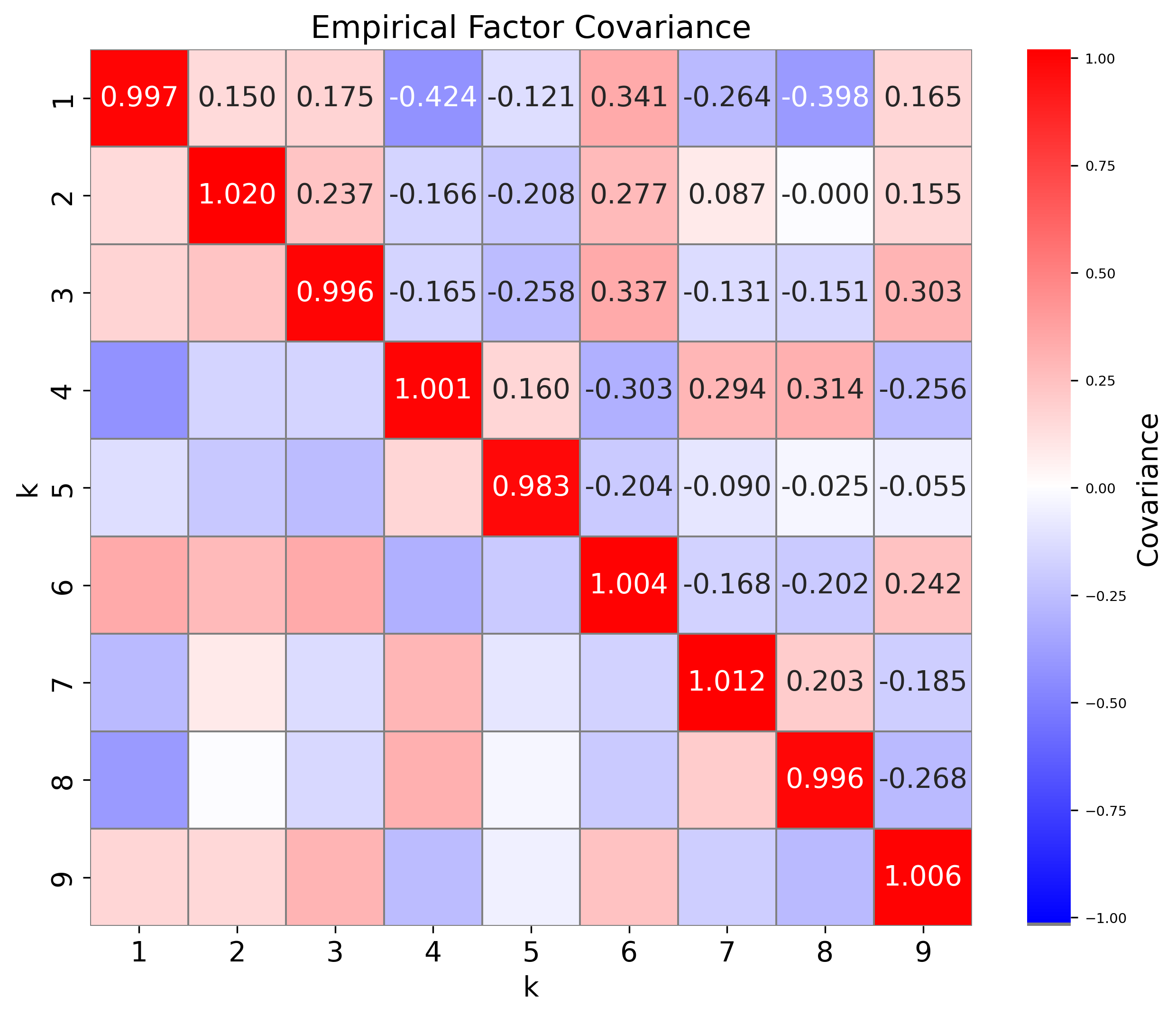}}
\caption{Diagnostic heatmaps displaying pairwise factor covariances for the 210-subject analysis. For the orthogonal diagnostic, an asterisk indicates that the covariance (or variance) differs significantly from zero (or one) at the 0.05 level without correcting for multiple testing. The two flagged comparisons are insignificant after application of a Bonferroni correction.}
\label{fig:aomic-210_fac-cov_ffa}
\end{figure}

This section presents an analysis of the full 210-subject dataset while Section \ref{sec:supp-aomic-20} of the supplement considers only the first 20 subjects, a size typical of small-scale fMRI studies. We begin analysis by constructing a mask that contains only the roughly $M = 54,000$ voxels that lie within the brain of all 210 subjects. Voxels outside this mask are henceforth ignored. We then compute only the entries of the empirical average spatial covariance $\hat{\mcal{C}}^s$ needed to fit the TFFM: entries for pairs of voxels that (i) lie within the brain mask, and (ii) are separated by a distance of $\delta = 0.1$ following transformation of volume dimensions to $[0,1]^3$. This covariance computation is performed in a batched and distributed manner to avoid memory limits and accelerate computation. Using the scree plot procedure, we then set the number of factors $\hat{K}$ equal to nine (see Figure \ref{fig:supp-aomic-210_rank} of the supplement for scree and ratio plots) and estimate loadings via our distributed algorithm, storing these estimates as a matrix $\hat{\mb{L}} \in \mbb{R}^{M \times \hat{K}}$. After separately applying both orthogonal (varimax) and oblique (quartimin) procedures, we smooth and shrink each column of the rotated matrices $\hat{\mb{L}}^*$ independently--choosing $\und{\sigma}$ and $\und{\kappa}$ via 3-fold cross-validation--to obtain the final postprocessed estimators $\bar{\mb{L}}$ (see Figure \ref{fig:aomic-210_ffa}). Lastly, we compute factor scores for both models via our F-on-S approach--tuning $\und{\gamma}$ via 4-fold spatial cross-validation--then estimate ``on average'' factor covariances which we visualize as heatmaps in Figure \ref{fig:aomic-210_fac-cov_ffa}.

Known resting-state networks are clearly present in the loading estimates of Figure \ref{fig:aomic-210_ffa}. For instance, much of the oft-studied ``default mode network'' (see $4_{20}$ of Figure 1 in \cite{smith-etal-2009}) may be found in the second empirical loading tensor of both models. Other well-known networks, like those associated with visual (see $1_{20}$, $2_{20}$, and $3_{20}$ of Figure 1 in \cite{smith-etal-2009}) and auditory ($7_{20}$) processing are present in the first and fourth loading estimates, respectively, of both models. 

Oblique rotation discovers noticeably simpler structures than does orthogonal rotation. For instance, the first loading in Figure \ref{fig:aomic-210_fac-cov_ffa-or} shares many of the structures found in the first loading of Figure \ref{fig:aomic-210_fac-cov_ffa-ob}, yet the latter has more regions that have been rotated towards zero. The same holds for many other loading pairs. This was expected as the much larger space of oblique rotations promotes discovery of simpler structures. In discovering such structure, the oblique rotation used in this analysis induced the factor correlations shown in Figure \ref{fig:aomic-210_fac-cov_ffa-ob}. Meanwhile, Figure \ref{fig:aomic-210_fac-cov_ffa-or} suggests that orthogonal rotation largely preserved uncorrelated factors. 

\begin{figure}[!h]
\centering
\includegraphics[width=0.35\linewidth]{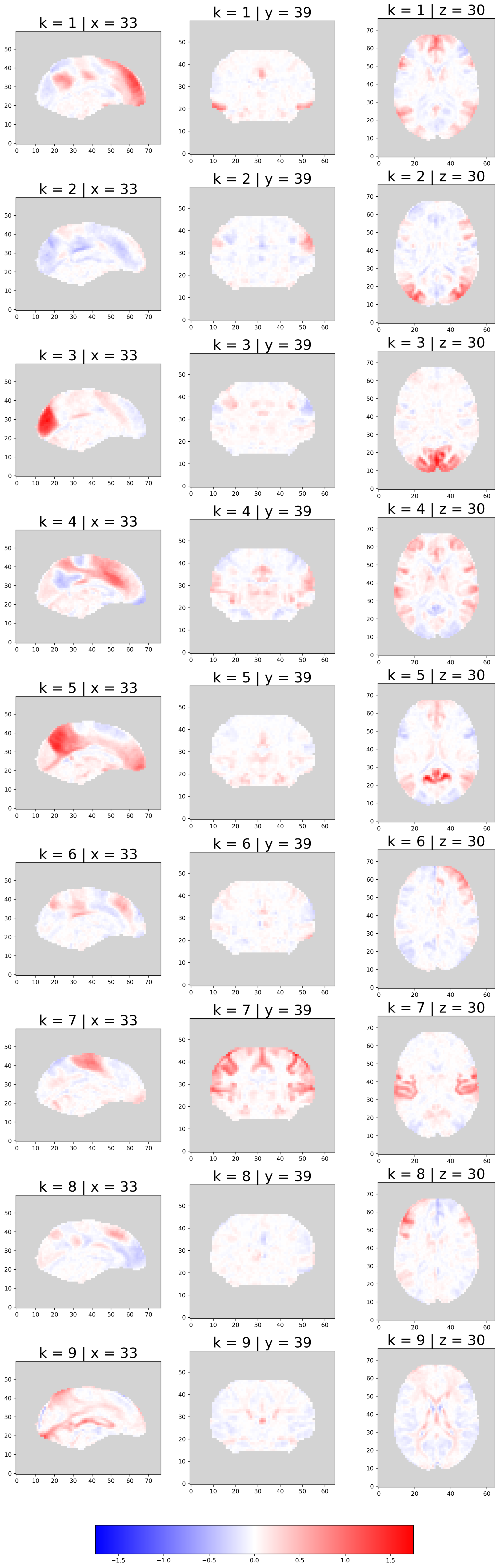}
\caption{IC estimates for the ICA model fit to data from 210 AOMIC subjects.}
\label{fig:aomic-210-ica}
\end{figure}

To provide a comparison, we present IC estimates for a 9-component ICA model fit to unsmoothed data from all 210 AOMIC subjects (see Figure \ref{fig:aomic-210-ica}). These estimates are remarkably similar to those of oblique TFFA, as was expected given our observations from the simulation study of Section \ref{sec:sim-s2} of the manuscript. Of course, the advantage of our framework is that, unlike ICA, we are free to entertain other representations of the target subspace. 

\section{Discussion}
\label{sec:disc}

This paper builds upon \cite{stanley-etal-2025}, delivering a comprehensive new approach to studying FC. We began by updating the model of \cite{stanley-etal-2025} to account for the temporal dependencies present in fMRI data. We then invoked the distributed LRCMC algorithm of \cite{stanley-etal-2025+} to scale estimation of the global spatial covariance to the full brain. After exploring the global subspace via factor rotation and postprocessing the rotated loadings via smoothing and shrinkage, we computed factor scores using F-on-S regression techniques. The result is TFFA, an LLVM-based approach that (1) defines the target subspace as describing all sustained global spatial variation, then (2) rotates an initial expression of this subspace to optimize a user-specified criterion.

Users of our approach should follow a couple of practical guidelines. First, data smoothing should not precede our approach as it does for other FC methods. Pre-smoothing would smear local variation into the global signal, thereby corrupting subsequent attempts to estimate that signal. Second, when the number of factors is known (or can be reliably estimated), we recommend using oblique rotations unless there is one wishes to study only uncorrelated factors. Oblique rotations tend to discover simpler structure and, because they are a superset of orthogonal rotations, can still recover orthogonal factors. 

Throughout this work, we have emphasized the utility of TFFA as an exploratory tool. By leveraging the vast suite of rotation algorithms developed for classical FA, TFFA enables investigation of its target subspace, and insights from such exploration may inform hypothesis-driven research. However, we placed less emphasis on TFFA as a dimension reduction tool, particularly one whose output could serve as input to other analyses. For example, one could use subject-wise factor scores as features in classification problems or as predictors in regression frameworks. The use of TFFA output in downstream analyses presents several potentially fruitful lines of research.

The main drawback of TFFA is that it relies on a massive voxel-by-voxel spatial covariance and thus requires access to a high-performance computing environment. While this may seem a reasonable prerequisite, its direct competitors--namely, ICA--are lightweight enough to run on personal computers. Lowering the computational overhead of TFFA may hasten its adoption among practitioners.


\pagebreak
\begin{center}
\textbf{\LARGE Supplement: Temporal Functional Factor Analysis of Brain Connectivity}
\end{center}
\setcounter{section}{0}
\setcounter{equation}{0}
\setcounter{figure}{0}
\setcounter{table}{0}
\makeatletter
\vspace{1em}

\renewcommand{\theequation}{\arabic{equation}A}
\renewcommand{\thefigure}{\arabic{figure}A}
\renewcommand{\thetable}{\arabic{table}A}

\section{Data Availability and Computer Code}
\label{sec:supp-data-code}

The links below may be used to access both the AOMIC-PIOP1 dataset which is publicly available on OpenNeuro and code for reproducing the results of this paper which is hosted on GitHub. 
\begin{itemize}
    \item[] Data: \url{https://openneuro.org/datasets/ds002785/versions/2.0.0}
    \item[] Code: \url{https://github.com/kylestanley1313/ffa-p2}
\end{itemize}

\section{Transformational Indeterminacy}
\label{sec:supp-ti}

In this section, we describe why the orthogonal and oblique TFFMs are identifiable only up to orthonormal and invertible transformations, respectively. We present explanations in the infinite-resolution setting, omitting parallel finite-resolution discussions since the details are similar. In what follows, we condense notation by defining
\begin{align*}
    \und{l}(\und{s}) & = [l_1(\und{s}), \dots, l_K(\und{s})], \text{ and} \\
    \und{F}(t) & = [F_1(t), \dots, F_K(t)].
\end{align*}
Explanations in both settings will make use of the following spectral decomposition:
\begin{equation}
\label{eqn:supp-ti-edecomp}
    g^s(\und{s}, \und{s}') = \sum_{k=1}^K \lambda_k \eta_k(\und{s}) \eta_k(\und{s}') = \und{\Tilde{\eta}}(\und{s})^T \und{\Tilde{\eta}}(\und{s}'),
\end{equation}
where $\Tilde{\eta}_k = \lambda_k^{1/2} \eta_k$ denotes the $k$th scaled eigenfunction of $g^s$. If the global term $Y(\und{s},t)$ may be expressed using these scaled eigenfunctions, 
\begin{equation*}
    Y(\und{s},t) = \sum_{k=1}^K \Tilde{\eta}_k(\und{s}) c_k(t),
\end{equation*}
then $g^s$ takes the form
\begin{align*}
    g^s(\und{s}, \und{s}') = \sum_k \left( \int \mbb{E}[c_k^2(t)] dt \right) \lambda_k \eta_k(\und{s}) \eta_k(\und{s}') + \sum_{k \neq j} \left( \int \mbb{E}[c_k(t) c_j(t)] dt \right) \lambda_k^{1/2} \lambda_j^{1/2} \eta_k(\und{s}) \eta_j(\und{s}').
\end{align*}
For (\ref{eqn:supp-ti-edecomp}) to hold, we require that
\begin{align*}
    \int \mbb{E}[\und{c}(t) \und{c}(t)^T]dt = \mb{I}.
\end{align*}
Hence, the scaled eigenfunctions $\Tilde{\eta}_k$ and the functional coefficients $c_k$ serve as the loadings and factors, respectively, in a valid orthogonal TFFM. We invoke this fact in our discussions of transformational ambiguities in Sections \ref{sec:supp-ti-or} and \ref{sec:supp-ti-ob}.

\subsection{The Orthogonal Model}
\label{sec:supp-ti-or}

Recall that the global spatial covariance in the orthogonal TFFM decomposes as
\begin{equation}
\label{eqn:supp-ti-scov-or}
    g_s(\und{s}, \und{s}') = \und{l}(\und{s})^T \und{l}(\und{s}').
\end{equation}
Upon examination of (\ref{eqn:supp-ti-edecomp}) and (\ref{eqn:supp-ti-scov-or}), it becomes clear that rotating the scaled eigenfunctions $\Tilde{\eta}_k$ by some orthonormal matrix $\mb{R}$ recovers the loading functions $l_k$: 
\begin{equation*}
    \und{l}(\und{s}) = \und{\Tilde{\eta}}(\und{s}) \mb{R}.
\end{equation*}
However, any rotation of valid loadings (e.g., the scaled eigenfunctions) are equivalent under the model when paired with a corresponding rotation of the factors. To understand this ambiguity, suppose $\und{l}(\und{s})$ and $\und{F}(t)$ jointly satisfy assumptions of the orthogonal TFFM. Let $\mb{R}$ be some $K$-by-$K$ orthonormal matrix, and define rotations of the loadings and factors as follows: $\und{l}^*(\und{s}) = \und{l}(\und{s})\mb{R}$ and $\und{F}^*(t) = \und{F}(t)\mb{R}$. 
First, rotation preserves the global component $Y$:
\begin{align*}
    \sum_{k=1}^K l_k^*(\und{s})F^*_k(t) & = \und{l}^*(\und{s})[\und{F}^*(t)]^T \\
    & = \und{l}(\und{s})\mb{R} [\und{f}(t)\mb{R}]^T \\
    & = \und{l}(\und{s}) \mb{R} \mb{R}^T [\und{F}(t)]^T \\
    & = \und{l}(\und{s}) [\und{F}(t)]^T \\
    & = Y(\und{s}, t)
\end{align*}
Second, the rotated factors are still, on average, uncorrelated with unit variance:
\begin{align*}
    \int \mbb{E} \left[ [\und{F}^*(t)]^T \und{F}^*(t) \right] dt = \mb{R}^T \left( \int \mbb{E} \left[ [\und{F}(t)]^T \und{F}(t) \right] dt \right) \mb{R} = \mb{R}^T \mb{I} \mb{R} = \mb{I}
\end{align*}
Therefore, the model is preserved under orthogonal rotation of the factors and loadings.

\subsection{The Oblique Model}
\label{sec:supp-ti-ob}

Recall that in the oblique TFFM, the global spatial covariance decomposes as 
\begin{equation}
\label{eqn:supp-ti-scov-ob}
    g_s(\und{s}, \und{s}') = \und{l}(\und{s})^T \mb{H} \und{l}(\und{s}').
\end{equation}
Let $\mb{H} = \mb{V} \mb{\Lambda} \mb{V}^T$ be the eigendecomposition of $\mb{H}$. It is easy to show that the $\und{\Tilde{l}} = \und{l} \mb{V} \mb{\Lambda}^{1/2}$ and $\und{\Tilde{F}} = \mb{\Lambda}^{-1/2} \mb{V}^T \und{F}$ comprise the loadings and factors, respectively, of an orthogonal TFFM. As before, one may recover the loadings $\Tilde{l}_k$ of this orthogonal model by some orthogonal rotation $\mb{R}$ of the scaled eigenfunctions: 
\begin{equation}
\label{eqn:supp-ti-ob-trans-1}
    \und{\Tilde{l}} = \und{\Tilde{\eta}} \mb{R}.
\end{equation}
The true loadings $l_k$ of the oblique model may then be obtained by reversing the orthogonalizing transformation: 
\begin{equation}
\label{eqn:supp-ti-ob-trans-2}
    \und{l} = \und{\Tilde{l}} \mb{\Lambda}^{-1/2} \mb{V}^T
\end{equation}
Combining (\ref{eqn:supp-ti-ob-trans-1}) and (\ref{eqn:supp-ti-ob-trans-2}) gives an invertible transformation $\mb{T} = \mb{V} \mb{\Lambda}^{1/2} \mb{R}^T$ whose inverse recovers the loadings from the scaled eigenfunctions: 
\begin{equation*}
    \und{l} = \und{\Tilde{\eta}} \mb{T}^{-1}.
\end{equation*}
It is easy to show that, given our constraint on the diagonal elements of $\mb{H}$, this transformation $\mb{T}$ satisfies the constraint $\text{diag}(\mb{T}\mb{T}^T) = \mb{I}$. As we note in the manuscript, the oblique model admits more ambiguity than the orthogonal one. Not only may $\mb{R}$ be replaced by any orthonormal matrix, but since $\mb{H}$ is unknown, we cannot use its eigendecomposition to fix the last two matrices in the expression for $\mb{T}$.

\section{Postprocessing Tuning Procedures}
\label{sec:supp-tune}

To postprocess the rotated loading estimates $\hat{\mcal{L}}_k^*$, we must select smoothing and shrinkage parameters: $\und{\sigma}$ and $\und{\kappa}$. This is accomplished by separately applying one round of cross-validation to choose $\und{\sigma}$ and another to choose $\und{\kappa}$. We recommend smoothing before shrinkage as this order generally leads to sparser estimates. In what follows, suppose the data $\mcal{X}_1, \dots , \mcal{X}_n$ have been partitioned into $V$ folds of (nearly) equal size, and understand that a quantity with $(v)$ (resp., $(-v)$) in its superscript denotes one computed using only data from fold $v$ (resp., all data except those in fold $v$). Throughout, we will also assume that all fold-wise loading estimates (e.g., $(\hat{\mcal{L}}_k^{(-v)})^*$) have been properly aligned with their full data counterparts (e.g., $\hat{\mcal{L}}_k^*$) so that component-wise tuning parameters are applied to the correct components across folds. In practice, one may need to use target rotations to achieve suitable alignment. Finally, we define the following error function
\begin{equation*}
    \text{Err}\left( v; \{ \mcal{L}_k \}_{k=1}^{\hat{K}} \right) = \norm{ \mcal{Z} \circ \left( \hat{\mcal{C}}^{(v)} - \sum_{k=1}^{\hat{K}} \mcal{L}_k \otimes \mcal{L}_k \right) }_F^2,
\end{equation*}
where we have dropped the $s$ from the empirical spatial covariance for fold $v$ to simplify notation.

\paragraph{Tuning $\und{\sigma}$ (orthogonal):}
Suppose that we have obtained the $\hat{\mcal{L}}_k^*$ by applying an orthogonal rotation $\mb{R} \in \mbb{R}^{\hat{K}\times \hat{K}}$ to the $\hat{\mcal{L}}_k$. We may use the following cross-validation procedure to choose $\und{\sigma}^*$:
\begin{enumerate}
    \item For each candidate value of $\und{\sigma}$:
    \begin{enumerate}
        \item For each $v$, shrink the $(\Tilde{\mcal{L}}_k^{(-v)})^*$ using $\und{\sigma}$ to obtain the $(\Tilde{\mcal{L}}_k^{(-v)})^*(\und{\sigma})$. 
        \item Compute $CV(\und{\sigma}) = \frac{1}{V} \sum_{v=1}^V \text{Err}\left( v; \{ (\Tilde{\mcal{L}}_k^{(-v)})^*(\und{\sigma})\}_{k=1}^{\hat{K}} \right)$
    \end{enumerate}
    \item Set $\und{\sigma}^* = \underset{\und{\sigma}}{\text{argmin}} \ CV(\und{\sigma})$.
\end{enumerate}

\paragraph{Tuning $\und{\sigma}$ (oblique):}
Suppose that we have obtained the $\hat{\mcal{L}}_k^*$ by applying an oblique rotation $\mb{T} \in \mbb{R}^{\hat{K}\times \hat{K}}$ to the $\hat{\mcal{L}}_k$. To choose $\und{\sigma}^*$, we apply the following cross-validation procedure:
\begin{enumerate}
    \item For each candidate value of $\und{\sigma}$:
    \begin{enumerate}
        \item For each $v$:
        \begin{enumerate}
            \item Shrink the $(\Tilde{\mcal{L}}_k^{(-v)})^*$ using $\und{\sigma}$ to obtain the $(\Tilde{\mcal{L}}_k^{(-v)})^*(\und{\sigma})$. 
            \item Obtain the $\Tilde{\mcal{L}}_k^{(-v)}(\und{\sigma})$ by applying the oblique rotation $\mb{T}^{-1}$ to the $(\Tilde{\mcal{L}}_k^{(-v)})^*(\und{\sigma})$. This effectively reverts the original oblique rotation, ensuring that the factors are (approximately) orthogonal prior to computing the global spatial covariance in (b).
        \end{enumerate}
        \item Compute $CV(\und{\sigma}) = \frac{1}{V} \sum_{v=1}^V \text{Err}\left( v; \{ \Tilde{\mcal{L}}_k^{(-v)}(\und{\sigma})\}_{k=1}^{\hat{K}} \right)$
    \end{enumerate}
    \item Set $\und{\sigma}^* = \underset{\und{\sigma}}{\text{argmin}} \ CV(\und{\sigma})$.
\end{enumerate}

\paragraph{Tuning $\und{\kappa}$ (orthogonal):}
Suppose that we have obtained the $\Tilde{\mcal{L}}_k^*$ by applying an orthogonal rotation $\mb{R} \in \mbb{R}^{\hat{K}\times \hat{K}}$ to the $\hat{\mcal{L}}_k$ then smoothing. We may use the following cross-validation procedure to choose $\und{\kappa}^*$:
\begin{enumerate}
    \item For each candidate value of $\und{\kappa}$:
    \begin{enumerate}
        \item For each $v$, shrink the $(\Tilde{\mcal{L}}_k^{(-v)})^*$ using $\und{\kappa}$ to obtain the $(\Tilde{\mcal{L}}_k^{(-v)})^*(\und{\kappa})$. 
        \item Compute $CV(\und{\kappa}) = \frac{1}{V} \sum_{v=1}^V \text{Err}\left( v; \{ (\Tilde{\mcal{L}}_k^{(-v)})^*(\und{\kappa})\}_{k=1}^{\hat{K}} \right)$
    \end{enumerate}
    \item Set $\und{\kappa}^* = \underset{\und{\kappa}}{\text{argmin}} \ CV(\und{\kappa})$.
\end{enumerate}

\paragraph{Tuning $\und{\kappa}$ (oblique):}
Suppose that we have obtained the $\Tilde{\mcal{L}}_k^*$ by applying an oblique rotation $\mb{T} \in \mbb{R}^{\hat{K}\times \hat{K}}$ to the $\hat{\mcal{L}}_k$ then smoothing. To choose $\und{\kappa}^*$, we apply the following cross-validation procedure:
\begin{enumerate}
    \item For each candidate value of $\und{\kappa}$:
    \begin{enumerate}
        \item For each $v$:
        \begin{enumerate}
            \item Shrink the $(\Tilde{\mcal{L}}_k^{(-v)})^*$ using $\und{\kappa}$ to obtain the $(\Tilde{\mcal{L}}_k^{(-v)})^*(\und{\kappa})$. 
            \item Obtain the $\Tilde{\mcal{L}}_k^{(-v)}(\und{\kappa})$ by applying the oblique rotation $\mb{T}^{-1}$ to the $(\Tilde{\mcal{L}}_k^{(-v)})^*(\und{\kappa})$. This effectively reverts the original oblique rotation, ensuring that the factors are (approximately) orthogonal prior to computing the global spatial covariance in (b).
        \end{enumerate}
        \item Compute $CV(\und{\kappa}) = \frac{1}{V} \sum_{v=1}^V \text{Err}\left( v; \{ \Tilde{\mcal{L}}_k^{(-v)}(\und{\kappa})\}_{k=1}^{\hat{K}} \right)$
    \end{enumerate}
    \item Set $\und{\kappa}^* = \underset{\und{\kappa}}{\text{argmin}} \ CV(\und{\kappa})$.
\end{enumerate}

\section{Factor Score Estimation}
\label{sec:supp-fse}

Recall from the manuscript that factor scores are estimated by fixing the postprocessed loadings $\bar{\mcal{L}}_k$ within the TFFM then invoking function-on-scalar (F-on-S) regression techniques. In this section, we derive the factor score estimator then present a technique for tuning its smoothness parameter. Throughout, we simplify notation by defining $\mb{L} \in \mbb{R}^{M \times \hat{K}}$ to be the matrix whose $k$th column is the vectorization of $\bar{\mcal{L}}_k$.

\subsection{Derivation of Factor Score Estimator}
\label{sec:supp-fse-est}

Let $\mb{X} \in \mbb{R}^{M \times J}$, $M = M_1\dots M_D$, be the matrix of voxel time courses whose $j$th column is given by the vectorization of the brain volume at time point $j$. We assume each $F_k(t)$ lies in the span of some truncated basis $\{ e_1(t), \dots , e_P(t)\}$: 
\begin{equation*}
    F_k(t) = \sum_{p=1}^P a_{kp} e_p(t).
\end{equation*}
Using matrix notation, we assume $\und{F}(t) = \mb{A}\und{e}(t)$ where $\und{e}(t) = [e_1(t), \dots , e_P(t)]^T$ and $\mb{A}\in \mb{A}^{\hat{K}\times P}$ is the matrix of basis coefficients given by $(\mb{A})_{kp} = a_{kp}$. Estimating $\und{F}(t)$ thus amounts to estimating $\mb{A}$, which we accomplish via penalized least squares:
\begin{equation}
\label{eqn:supp-est-fse-a-1}
\underset{\mb{A} \in \mbb{R}^{\hat{K}\times P}}{\arg \min} \left\{  \frac{1}{2}\norm{\mb{X} - \mb{L} \mb{A} \mb{E}}_F^2 
 + \gamma \ip{\mb{A}^T \mb{A}}{\mb{D}}_{F} \right\}
\end{equation}
where $\mb{E} \in \mbb{R}^{P \times J}$ is the matrix of discretized basis functions given by $(\mb{E})_{pj} = e_p(t_j)$, $\mb{D} \in \mbb{R}^{P \times P}$ is the matrix of second derivative inner products given by $(\mb{D})_{p_1,p_2} = \ip{e_{p_1}^{''}}{e_{p_2}^{''}}_2$, and $\gamma > 0$ is a tuning parameter. The first term in the objective of (\ref{eqn:supp-est-fse-a-1}) quantifies the distance between the observations and global components, while the second penalizes roughness in the $F_k$. Differentiating both terms with respect to $\mb{A}$ gives
\begin{align*}
    \frac{\partial \mscr{L}}{\partial \mb{A}} & = - \mb{L}^T \mb{X} \mb{E}^T + \mb{L}^T \mb{L} \mb{A} \mb{E} \mb{E}^T, \text{ and} \\
    \frac{\partial \mscr{P}}{\partial \mb{A}} & = \mb{A} \mb{D}.
\end{align*}
The $\mb{A}$ minimizing (\ref{eqn:supp-est-fse-a-1}) is thus the solution to 
\begin{align*}
    0 = - \mb{L}^T \mb{X} \mb{E}^T + \mb{L}^T \mb{L} \mb{A} \mb{E} \mb{E}^T + \gamma \mb{I} \mb{A} \mb{D}.
\end{align*}
After rearranging terms, invoking the identity 
\begin{align*}
    \text{vec}(\mb{A}_1 \mb{B} \mb{C}_1^T + \mb{A}_2 \mb{B} \mb{C}_2^T) = (\mb{C}_1 \otimes \mb{A}_1 + \mb{C}_2 \otimes \mb{A}_2) \text{vec}(\mb{B}),
\end{align*}
then rearranging terms once more, one arrives at the estimate of $\mb{A}$: 
\begin{equation}
\label{eqn:est-fse-a-2}
    \text{vec}(\hat{\mb{A}}) = \left( \mb{E}\mb{E}^T \otimes (\bar{\mb{L}})^T \bar{\mb{L}} + \mb{D} \otimes \gamma \mb{I}\right)^{-1} \text{vec}\left((\bar{\mb{L}})^T \mb{X} \mb{E}^T\right).
\end{equation}

\subsection{Tuning $\gamma$}
\label{sec:supp-fse-tune}

The estimator for $\mb{A}$ depends on a smoothing parameter $\gamma$ (or $\und{\gamma}$ for non-uniform smoothness) that must be tuned. To accomplish this, we use a spatial $V$-fold cross-validation procedure that selects an optimal parameter $\gamma^*$ from a collection of candidates $\{ \gamma_n \}_{n=1}^N$. The basic idea is to create (nearly) independent folds by blocking space. 
\begin{enumerate}
    \item Partition the rows of $\mb{X} \in \mbb{R}^{M \times J}$ (resp., $\mb{L} \in \mbb{R}^{M\times \hat{K}}$) into $V$ folds of (roughly) equal size so that the rows in each fold correspond to voxels that are spatially contiguous. Let $\mb{X}^{(v)} \in \mbb{R}^{M_v \times J}$ (resp., $\mb{L}^{(v)} \in \mbb{R}^{M_v \times \hat{K}}$) be the data (resp., loading) matrix for the $v$th fold, and $\mb{X}^{(-v)} \in \mbb{R}^{(M - M_v) \times J}$ (resp., $\mb{L}^{(-v)} \in \mbb{R}^{(M-M_v) \times \hat{K}}$) be the data (resp., loading) matrix for all but the $v$th fold. 
    \item For $n = 1, \dots, N$:
    \begin{enumerate}
        \item For $v = 1, \dots, V$:
        \begin{enumerate}
            \item Estimate factor scores using all but the $v$th fold:  $\hat{\mb{F}}^{(j,-v)} = \hat{\mb{A}}^{(j,-v)} \mb{E}$ where 
            \begin{equation*}
                \text{vec}(\hat{\mb{A}}^{(j,-v)}) = \left(\mb{E}\mb{E}^T \otimes \left(\mb{L}^{(-v)}\right)^T \mb{L}^{(-v)} + \mb{D} \otimes \gamma_j \mb{I} \right)^{-1} \text{vec} \left(  \left(\mb{L}^{(-v)}\right)^T \mb{X}^{(-v)} \mb{E}^T\right).
            \end{equation*}
            \item Compute the reconstruction error for the $v$th fold using $\hat{\mb{F}}^{(j,-v)}$:
            \begin{equation*}
                e_{jv} = M_v^{-1} \norm{\mb{X}^{(v)} - \mb{L}^{(v)} \hat{\mb{F}}^{(j,-v)}}_F^2.
            \end{equation*}
        \end{enumerate}
        \item Compute $e_j = V^{-1} \sum_v e_{jv}$.
    \end{enumerate}
    \item Set $\gamma^*$ equal to the $\gamma_j$ furnishing the minimum $E_j$.
\end{enumerate}
Throughout the manuscript, we tune $\gamma$ using four-fold spatial cross-validation. When the data are comprised of spatial ``volumes'' that take values in $[0,1]^2$ (as in our simulation study) these four folds are demarcated by the lines $x = 1/2$ and $y = 1/2$. When volumes take values in $[0,1]^3$ (as in our data analysis) the four folds are delimited by the planes $x = 1/2$ and $y = 1/2$. 

\section{Simulation Studies}
\label{sec:supp-sim}

This section presents supplementary material from the simulation study of the manuscript. 

\subsection{Study 2: Global Covariance Expression Under Dimension Overspecification}
\label{sec:supp-sim-s2}

\begin{figure}[!h]
    \centering
    \includegraphics[width=0.8\linewidth, trim=0 150 0 150, clip]{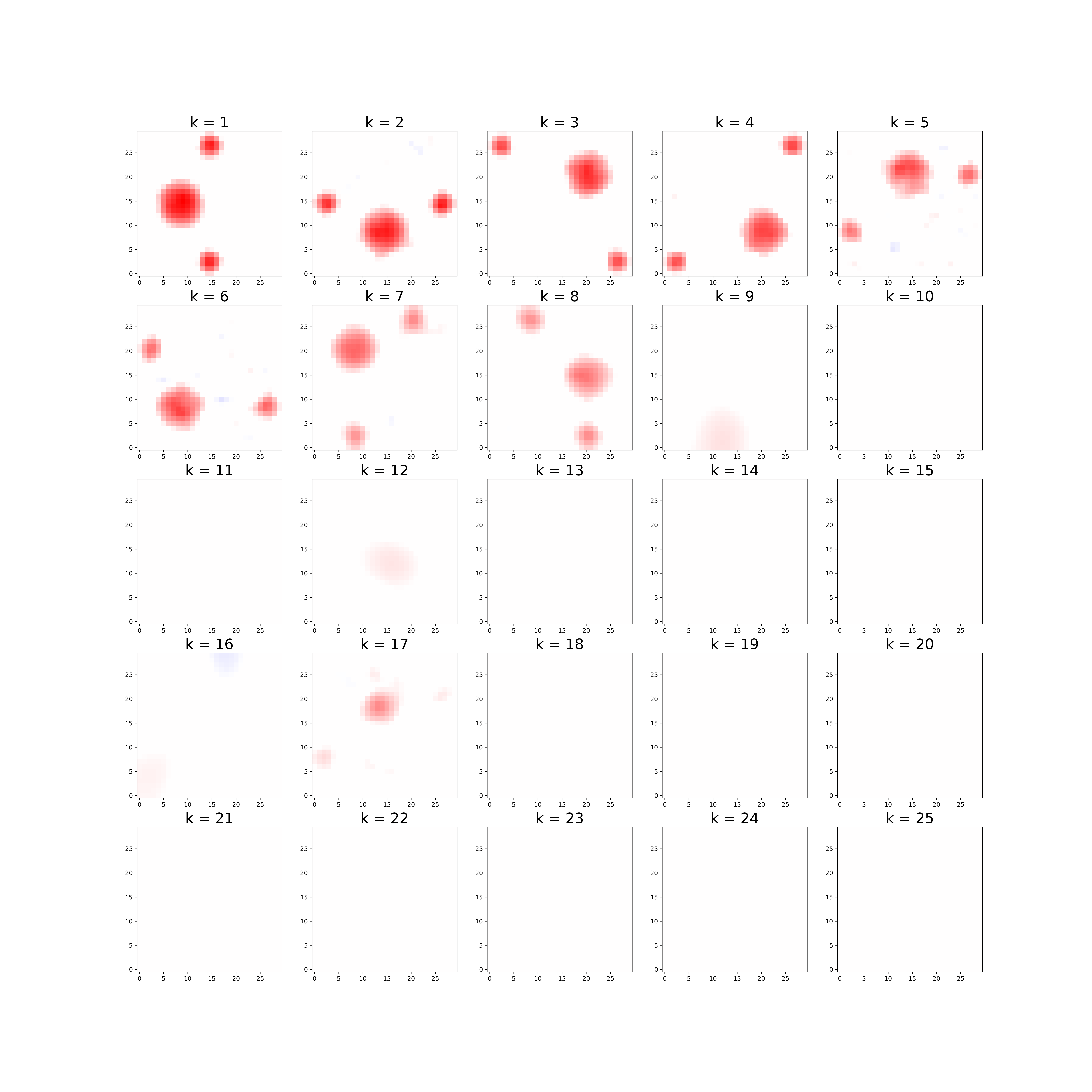}
    \caption{Estimates for the order-25 orthogonal TFFA model when data are generated from an orthogonal model in scenario of Study 2.}
    \label{fig:supp-sim-exp-25-or_ffa-or}
\end{figure}

\begin{figure}[!h]
    \centering
    \includegraphics[width=0.8\linewidth, trim=0 150 0 150, clip]{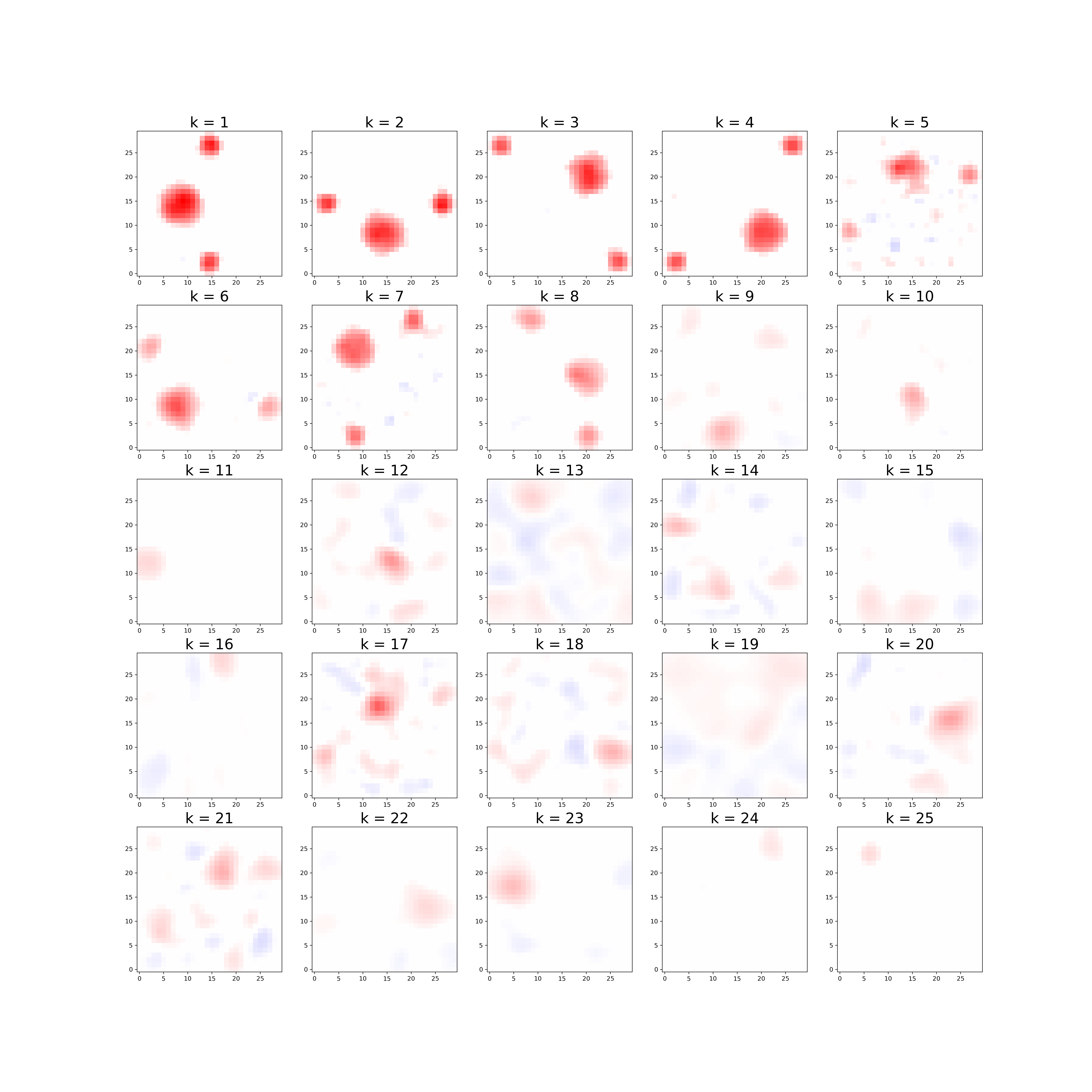}
    \caption{Estimates for the order-25 oblique TFFA model when data are generated from an orthogonal model in scenario of Study 2.}
    \label{fig:supp-sim-exp-25-or_ffa-ob}
\end{figure}

\begin{figure}[!h]
    \centering
    \includegraphics[width=0.8\linewidth, trim=0 150 0 150, clip]{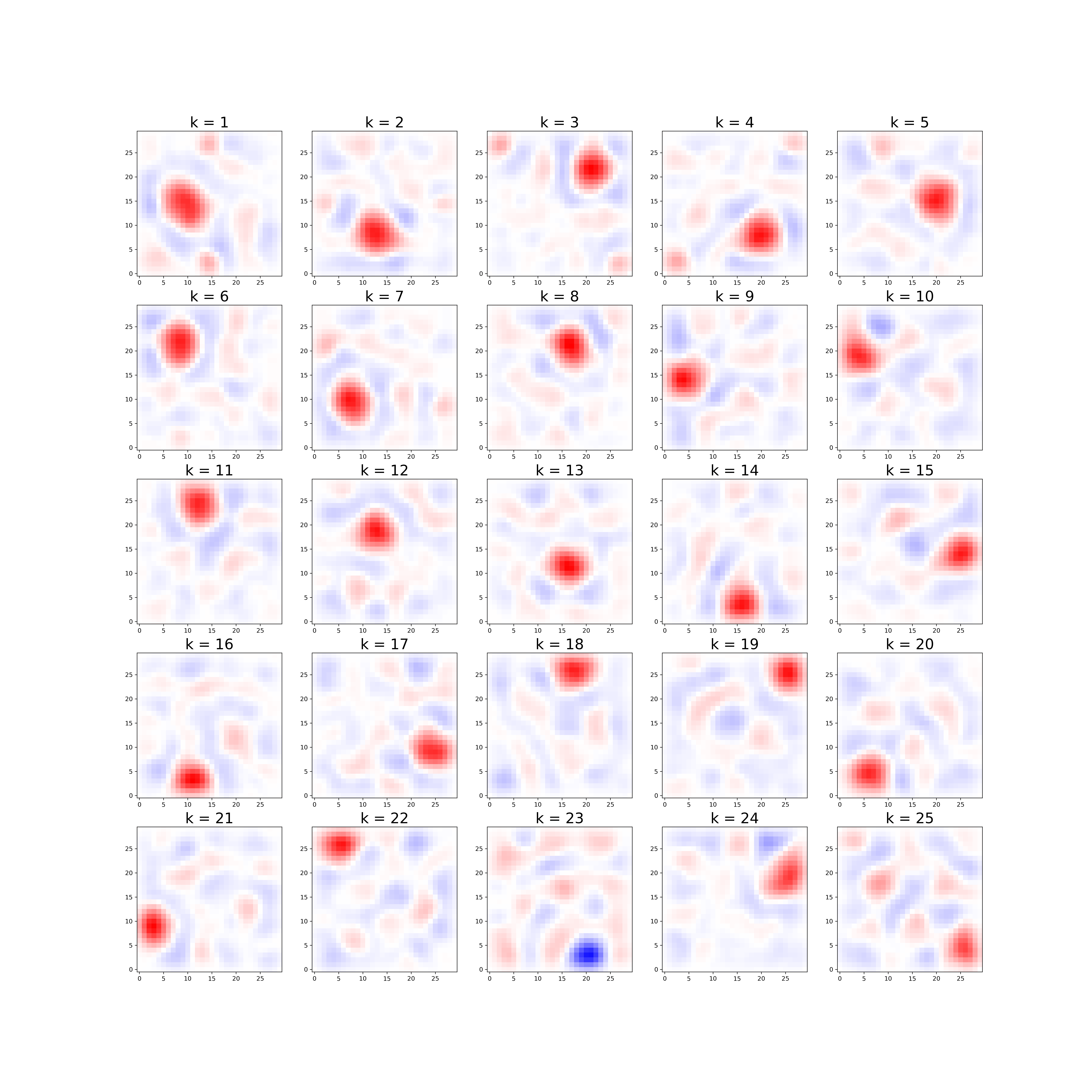}
    \caption{Estimates for the order-25 ICAS model when data are generated from an orthogonal model in scenario of Study 2.}
    \label{fig:supp-sim-exp-25-or_icas}
\end{figure}

In this section, we study subspace expression when model dimension is overspecified (i.e., $\hat{K} > K$). As in Section \ref{sec:sim-s2} of the manuscript, we consider regime-two scenarios with loading scheme set to TRI, $K$ set to 8, $n$ set to 20, and $\delta$ set to $0.1$. This time, however, we set $\hat{K} = 25$. Figures \ref{fig:supp-sim-exp-25-or_ffa-or}, \ref{fig:supp-sim-exp-25-or_ffa-ob}, and \ref{fig:supp-sim-exp-25-or_icas} present loading estimates for orthogonal TFFA, oblique TFFA, and ICAS, respectively, when data are generated from an orthogonal TFFM. The loadings estimated via orthogonal TFFA are most closely aligned to the truth, and include many trailing loadings that have been shrunk to zero. Although oblique TFFA captures much ground truth structure, its presentation is messier than that of orthogonal TFFA. This is because, in pursuit of simple structure, the more flexible oblique rotation algorithm splinters single-factor signals across multiple loadings (e.g., $k = 5,17,21$ and $k = 6,14$). Moreover, subsequent postprocessing more cautiously smooths and shrinks trailing loadings that contain these signal fragments, which prevents the convenient zeroing out achieved by orthogonal TFFA. Thankfully, it is rather easy to spot this fragmentation of global structure since all non-global structure has been eliminated. The same cannot be said of ICAS which splinters global structure among local structure (which is included in the target subspace), resulting in an indecipherable soup of global-local variation.

\begin{figure}[!h]
    \centering
    \includegraphics[width=0.8\linewidth, trim=0 150 0 150, clip]{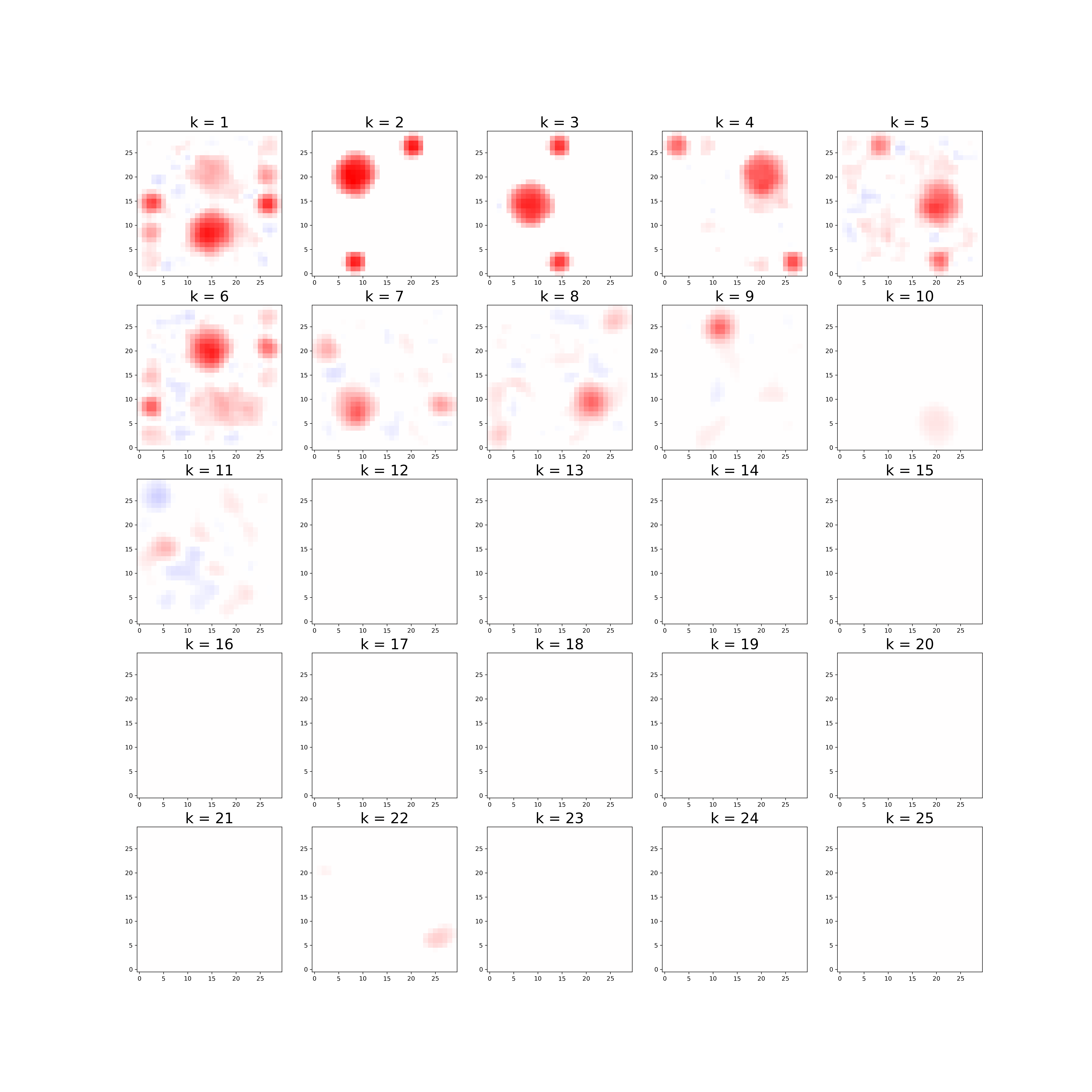}
    \caption{Estimates for the order-25 orthogonal TFFA model when data are generated from an oblique model in scenario of Study 2.}
    \label{fig:supp-sim-exp-25-ob_ffa-or}
\end{figure}

\begin{figure}[!h]
    \centering
    \includegraphics[width=0.8\linewidth, trim=0 150 0 150, clip]{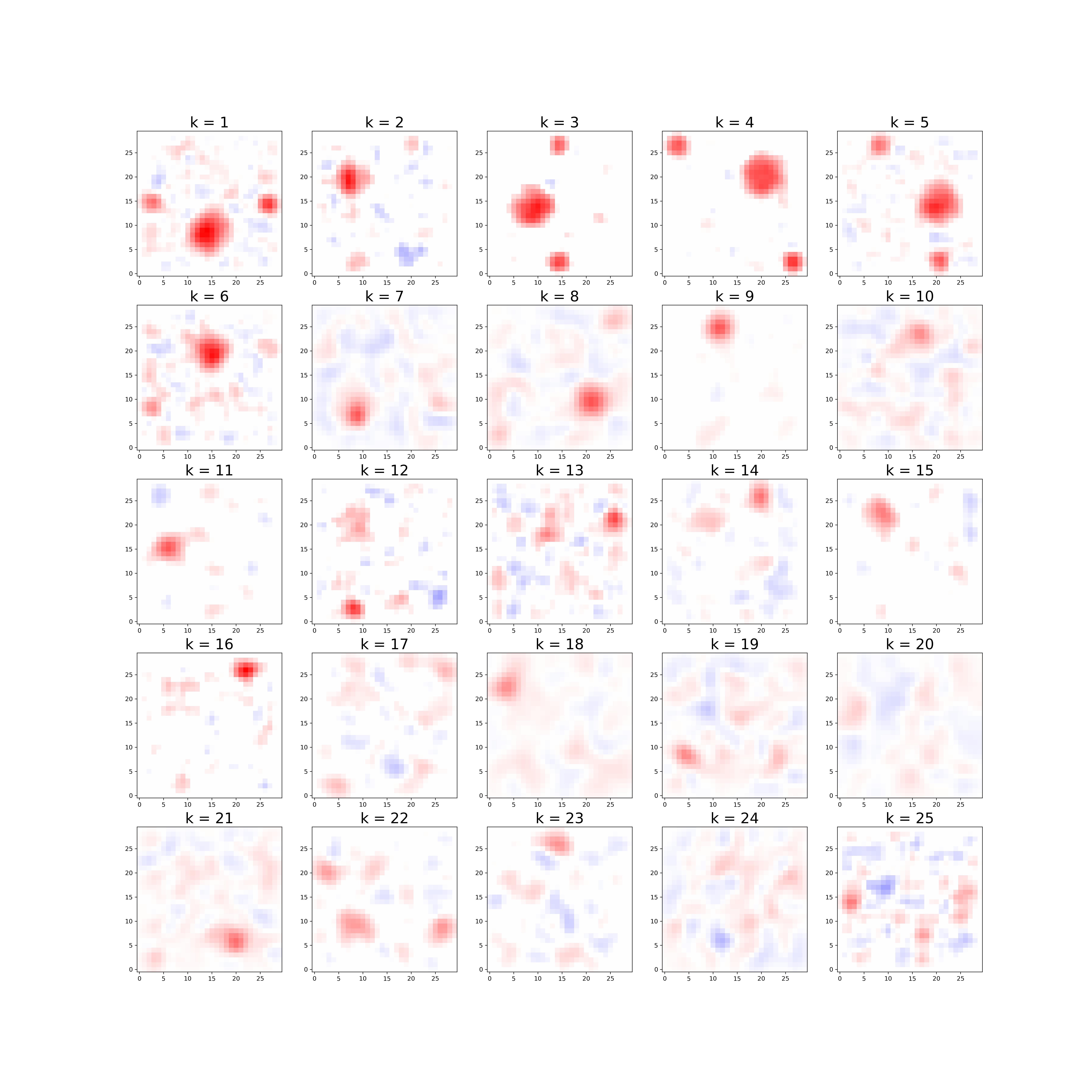}
    \caption{Estimates for the order-25 oblique TFFA model when data are generated from an oblique model in scenario of Study 2.}
    \label{fig:supp-sim-exp-25-ob_ffa-ob}
\end{figure}

\begin{figure}[!h]
    \centering
    \includegraphics[width=0.8\linewidth, trim=0 150 0 150, clip]{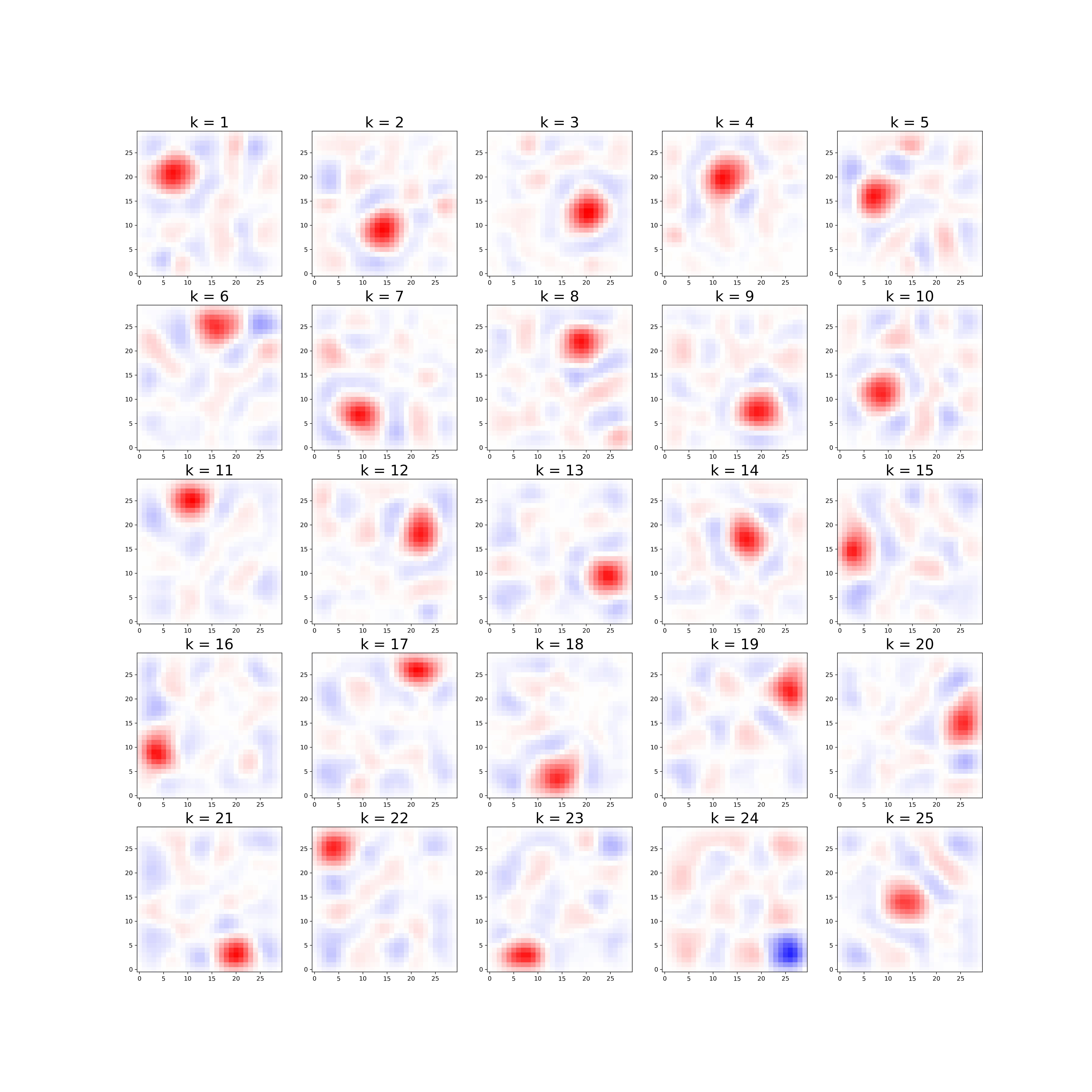}
    \caption{Estimates for the order-25 ICAS model when data are generated from an oblique model in scenario of Study 2.}
    \label{fig:supp-sim-exp-25-ob_icas}
\end{figure}

The takeaways are similar when the data are generated from an oblique TFFM (see Figures \ref{fig:supp-sim-exp-25-ob_ffa-or}, \ref{fig:supp-sim-exp-25-ob_ffa-ob}, and \ref{fig:supp-sim-exp-25-ob_icas} for loadings estimated by orthogonal TFFA, oblique TFFA, and ICAS, respectively). The main difference is that, as when $K$ is correctly specified, orthogonal TFFA sometimes combines signals from multiple factors into a single loading (e.g., $k = 1$).  

This study suggests that orthogonal TFFA is robust to dimension overspecification when the factors are, in fact, uncorrelated. However, if (i) factors are correlated and/or (ii) one uses an oblique TFFA, then model dimension should be chosen carefully.

\clearpage

\section{AOMIC Analysis}
\label{sec:supp-aomic}

This section contains spillover content from the 210-subject analysis of Section \ref{sec:aomic} of the manuscript, as well as a 20-subject analysis whose size is closer to that of small-scale FC studies.

\subsection{210-Subject Analysis}
\label{sec:supp-aomic-210}

\begin{figure}[!h]
\centering
{\label{fig:supp-aomic-210_scree}\includegraphics[width=0.35\linewidth]{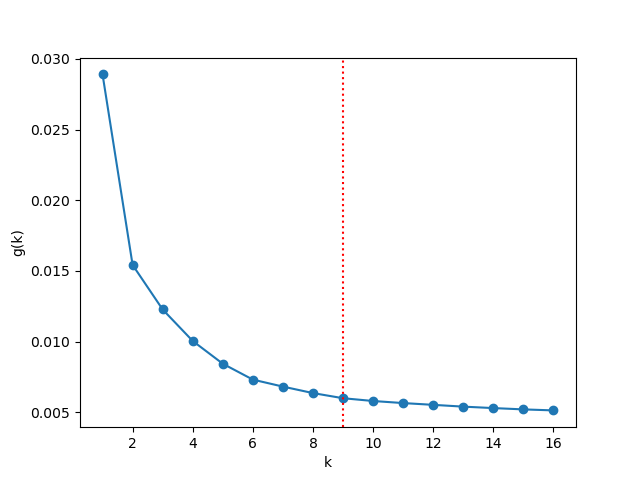}}\qquad
{\label{fig:supp-aomic-210_ratio}\includegraphics[width=0.35\linewidth]{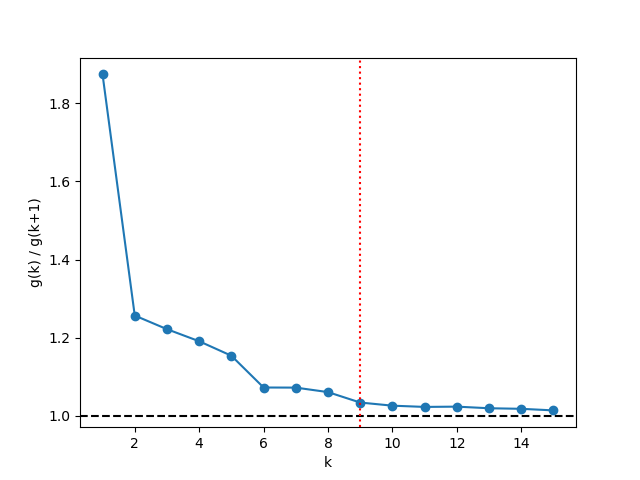}}
\caption{The scree (left) and ratio (right) plot used to select the number of factors in the 210-subject TFFMs.}
\label{fig:supp-aomic-210_rank}
\end{figure}

In the 210-subject analysis, we set the number of factors equal to nine, guided by the scree and ratio plots of Figure \ref{fig:supp-aomic-210_rank}.

\subsection{20-Subject Analysis}
\label{sec:supp-aomic-20}

In this Section, we use TFFA to study resting-state data from the first 20 AOMIC subjects. Many results from this 20-subject analysis are in close alignment with those of the 210-subject analysis. We exclude discussion of these similar findings to avoid redundancy, but highlight any noteworthy differences between the two analyses. 

\begin{figure}[!h]
\centering
{\label{fig:supp-aomic-20_scree}\includegraphics[width=0.35\linewidth]{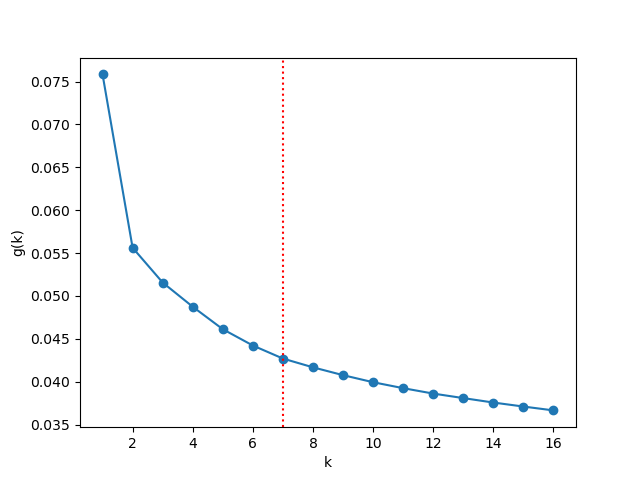}}\qquad
{\label{fig:supp-aomic-20_ratio}\includegraphics[width=0.35\linewidth]{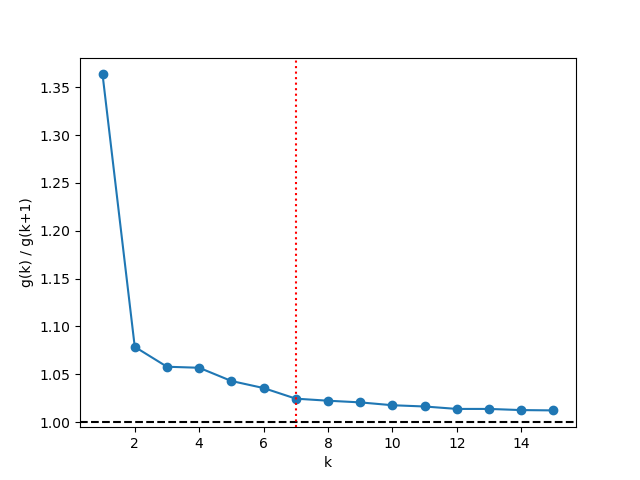}}
\caption{The scree (left) and ratio (right) plot used to select the number of factors in the 20-subject TFFMs.}
\label{fig:supp-aomic-20_rank}
\end{figure}

\begin{figure}[!h]
\centering
\subfloat[Orthogonal TFFA]{\label{fig:supp-aomic-20_ffa-or}\includegraphics[width=0.35\linewidth]{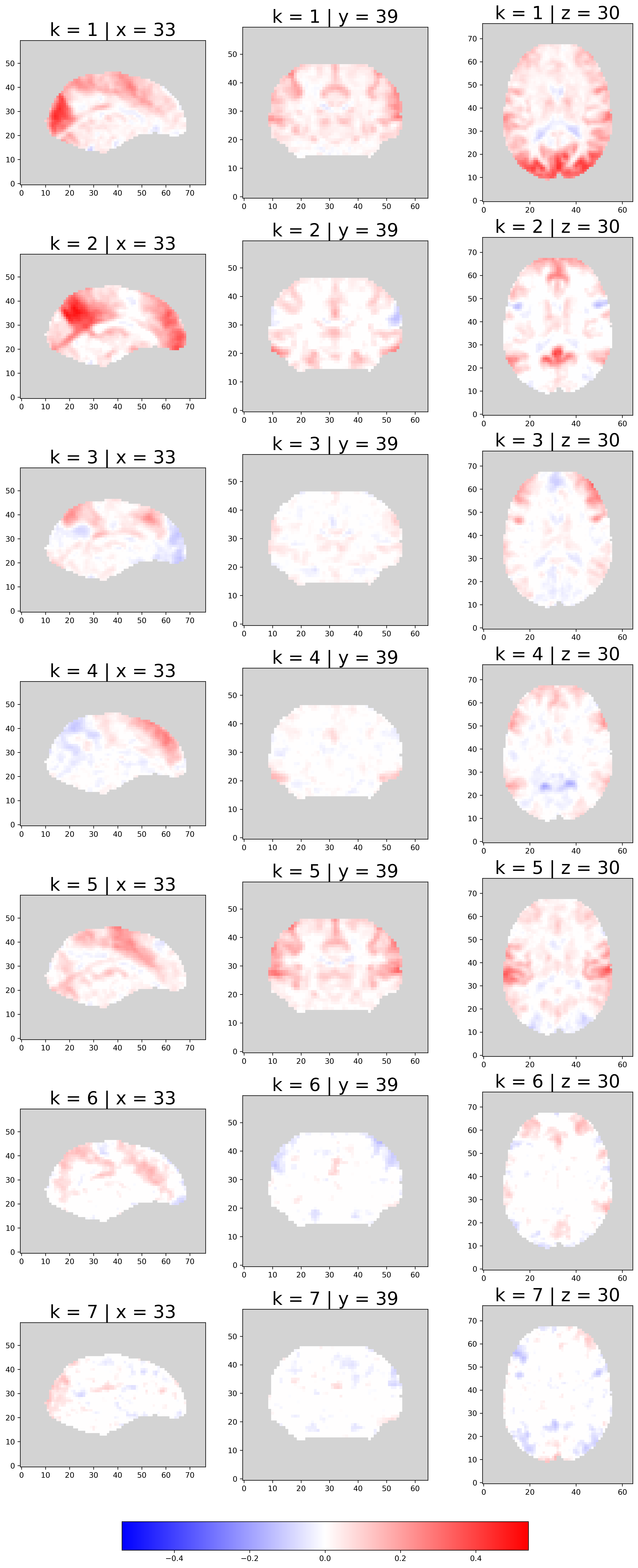}}\qquad
\subfloat[Oblique TFFA]{\label{fig:supp-aomic-20_ffa-ob}\includegraphics[width=0.35\linewidth]{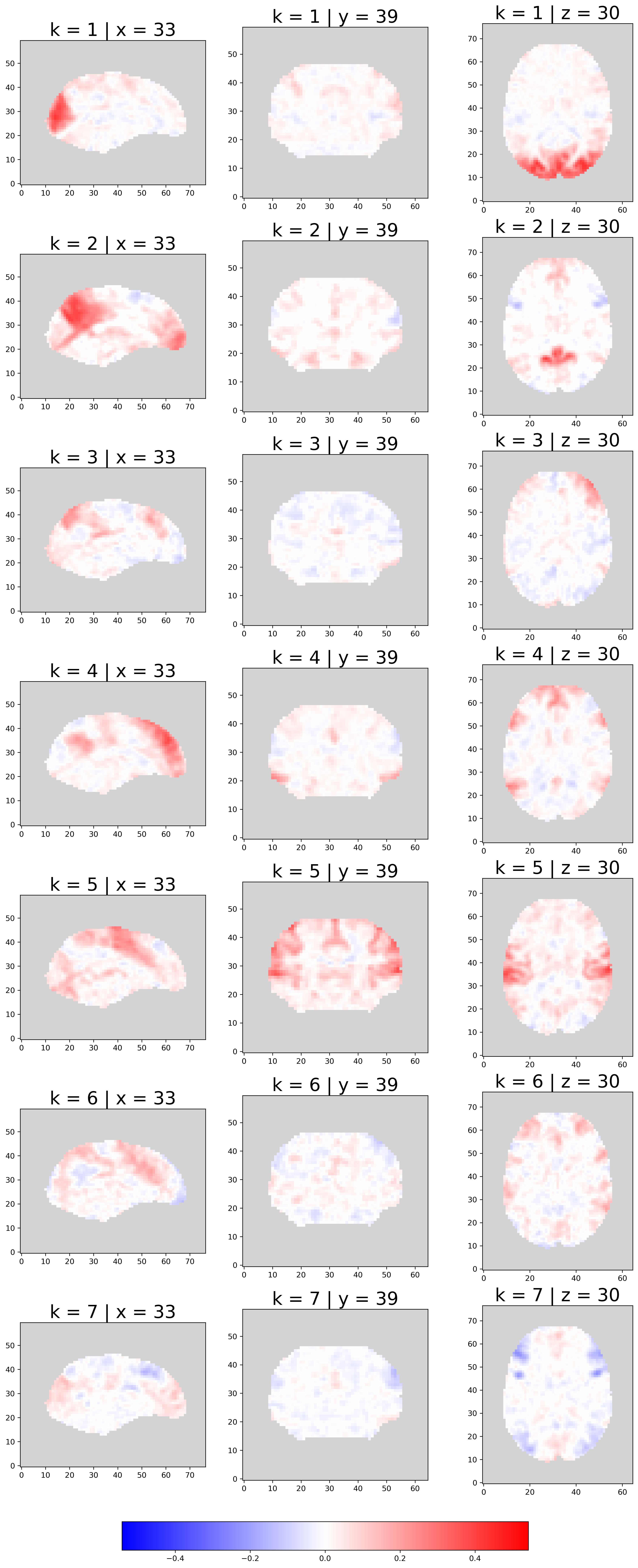}}
\caption{Loading estimates for the orthogonal and oblique TFFMs fit to data from 20 AOMIC subjects.}
\label{fig:supp-aomic-20_ffa}
\end{figure}

After setting the number of factors to seven using the scree-plot approach (see Figure \ref{fig:supp-aomic-20_rank}), we estimate and postprocess loadings for the orthogonal and oblique models (see Figure \ref{fig:supp-aomic-20_ffa}). Though these loadings contain much of the structure found in the 210-subject loadings, smoothing and (especially) shrinkage clearly has a larger impact in the smaller study. This is, of course, not surprising. In most estimation problems, the effects of regularization are stronger as sample size decreases.

\begin{figure}[!h]
\centering
\subfloat[Orthogonal TFFA]{\label{fig:supp-aomic-20_fac-cov_ffa-or}\includegraphics[width=0.45\linewidth]{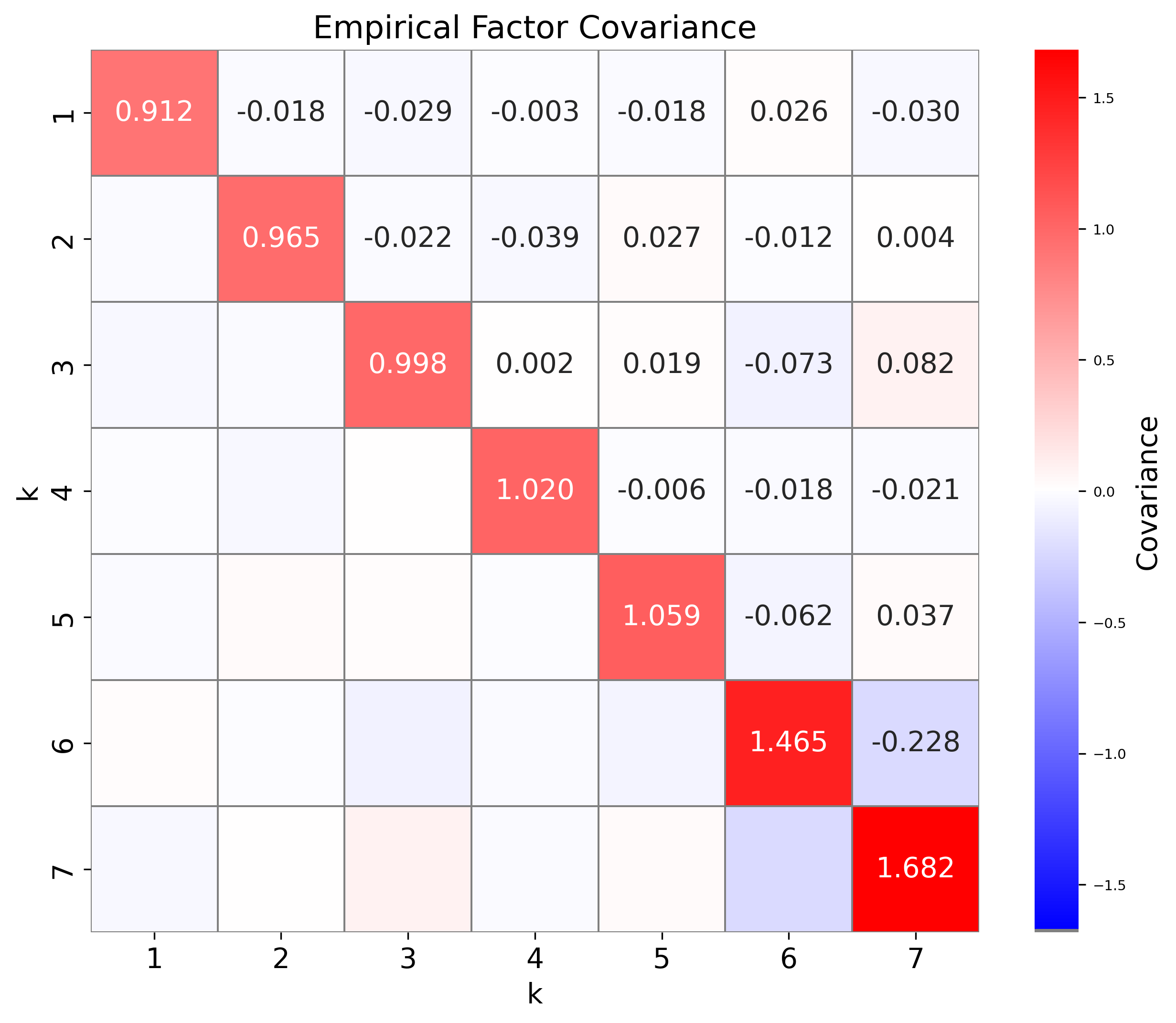}}\qquad
\subfloat[Oblique TFFA]{\label{fig:supp-aomic-20_fac-cov_ffa-ob}\includegraphics[width=0.45\linewidth]{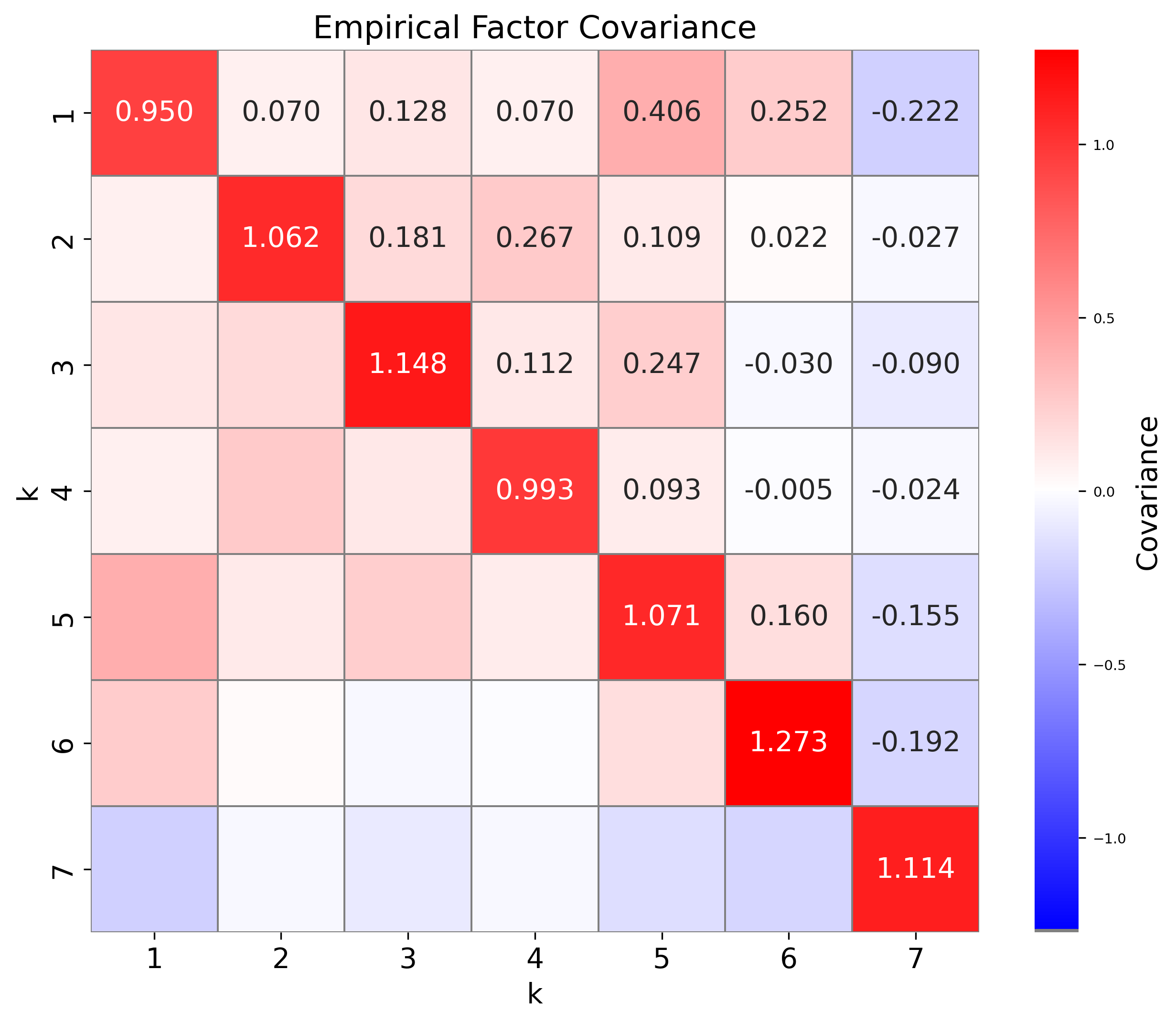}}
\caption{Diagnostic heatmaps displaying pairwise factor covariances for the 20-subject analysis. For the orthogonal diagnostic, an asterisk (of which there are none) indicates that the covariance (or variance) differs significantly from zero (or one) at the 0.05 level after correcting for multiple testing via a Bonferroni adjustment.}
\label{fig:supp-aomic-20_fac-cov_ffa}
\end{figure}

After obtaining the postprocessed loadings of Figure \ref{fig:supp-aomic-20_ffa}, we compute factor scores for both models using the F-on-S regression method outlined in Section \ref{sec:est-fse} of the manuscript. The heatmaps of Figure \ref{fig:supp-aomic-20_fac-cov_ffa} depict empirical factor covariances for both models. As in the 210-subject study, oblique rotation clearly induces correlation among the factors. However, the 20-subject orthogonal covariance seems to deviate more from the identity than than that of the 210-subject study (see Figure \ref{fig:aomic-210_fac-cov_ffa-or} of the manuscript): several off-diagonal elements are quite far from zero and the last diagonal element is nearly two. Yet the lack of asterisks in the heatmap of Figure \ref{fig:supp-aomic-20_fac-cov_ffa-or} indicate that the orthogonal covariance does not, in fact, differ significantly from the identity. This initially counterintuitive result arises from the fact that there are fewer subjects in this analysis, leading to higher-variance empirical factor covariances. Practitioners should be mindful of the diminished sensitivity of diagnostic testing in such small studies.

\begin{figure}[!h]
\centering
\includegraphics[width=0.35\linewidth]{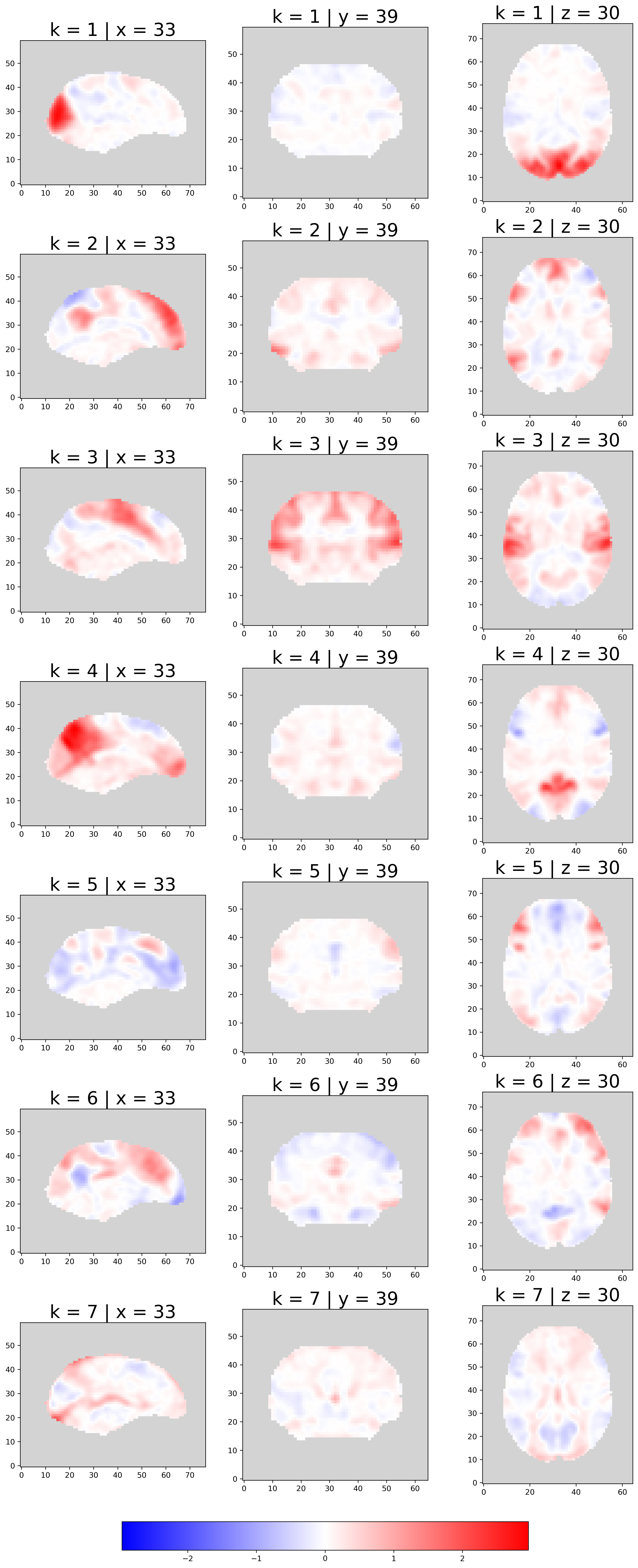}
\caption{IC estimates for the ICA model fit to data from 20 AOMIC subjects.}
\label{fig:supp-aomic-20-ica}
\end{figure}

As in the 210-subject study, we provide a comparison to TFFA by fitting a 7-component ICA model to the 20-subject data after smoothing each scan with a 1 mm Gaussian filter (see Figure \ref{fig:supp-aomic-20-ica}). The takeaway is similar to that of the 210-subject analysis: ICA estimates are similar to those of oblique TFFA, but TFFA allows users to more freely explore its target subspace.

\clearpage

\bibliographystyle{plainnat}
\bibliography{references}

\end{document}

%% file: custom_commands.tex
\usepackage{amsthm}
\usepackage{amsmath}
\usepackage{amssymb}
\usepackage{commath}
\usepackage{mathtools}
\usepackage{mathrsfs}
\usepackage[dvipsnames]{xcolor}

\usepackage{dsfont}

\newcommand{\und}[1]{\underline{#1}}
\newcommand{\mb}[1]{\boldsymbol{#1}}
\newcommand{\mbb}[1]{\mathbb{#1}}
\newcommand{\mcal}[1]{\mathcal{#1}}
\newcommand{\mscr}[1]{\mathscr{#1}}
\newcommand{\mfrak}[1]{\mathfrak{#1}}
\newcommand{\ip}[2]{\left\langle #1 , #2 \right\rangle}

\newtheorem{remark}{Remark}[section]